\documentclass[prd,superscriptaddress,amsfonts,amssymb,amsmath,showpacs,twocolumn]{revtex4-2}
\usepackage{bm}
\usepackage{amsfonts}
\usepackage{latexsym}
\usepackage[latin1]{inputenc}
\usepackage{graphicx}
\usepackage{amsmath}
\usepackage{palatino}
\usepackage{mathpazo}
\usepackage{textcomp}
\usepackage{float}
\usepackage{booktabs}
\usepackage{dcolumn}
\usepackage{ragged2e}
\usepackage{hyperref}
\hypersetup{colorlinks,citecolor=blue}
\hypersetup{colorlinks=true,linkcolor=red,filecolor=magenta,    urlcolor=blue}
\usepackage{amsmath}
\usepackage{xcolor}
\usepackage{orcidlink}
\usepackage[caption=false]{subfig}
\usepackage{commath}
\def\jnl@style{\it}
\def\aaref@jnl#1{{\jnl@style#1}}

\def\aaref@jnl#1{{\jnl@style#1}}

\def\aj{\aaref@jnl{AJ}}                   
\def\apj{\aaref@jnl{ApJ}}                 
\def\apjl{\aaref@jnl{ApJ}}                
\def\apjs{\aaref@jnl{ApJS}}               
\def\apss{\aaref@jnl{Ap\&SS}}             
\def\aap{\aaref@jnl{A\&A}}                
\def\aapr{\aaref@jnl{A\&A~Rev.}}          
\def\aaps{\aaref@jnl{A\&AS}}              
\def\mnras{\aaref@jnl{Mon.~Not.~Roy.~Astron.~Soc.}}             
\def\prd{\aaref@jnl{Phys.~Rev.~D}}        
\def\plb{\aaref@jnl{Phys.~Lett.~B}}        
\def\prc{\aaref@jnl{Phys.~Rev.~C}}  
\def\prl{\aaref@jnl{Phys.~Rev.~Lett.}}    
\def\qjras{\aaref@jnl{QJRAS}}             
\def\skytel{\aaref@jnl{S\&T}}             
\def\ssr{\aaref@jnl{Space~Sci.~Rev.}}     
\def\zap{\aaref@jnl{ZAp}}                 
\def\nat{\aaref@jnl{Nature}}              
\def\aplett{\aaref@jnl{Astrophys.~Lett.}} 
\def\apspr{\aaref@jnl{Astrophys.~Space~Phys.~Res.}} 
\def\physrep{\aaref@jnl{Phys.~Rep.}}      
\def\physscr{\aaref@jnl{Phys.~Scr}}       
\def\commat{\aaref@jnl{Comm.~Math.~Phys.}}              
\def\science{\aaref@jnl{Science}}               
\def\cqg{\aaref@jnl{Classical Quant.~Grav.}}            
\def\jpcs{\aaref@jnl{JPCS}}                                     
\def\ijmpd{\aaref@jnl{Int.~J.~Mod.~Phys.~D}}                    
\def\grg{\aaref@jnl{Gen.~Relat.~Gravit.}}               
\def\rpp{\aaref@jnl{Rep.~Prog.~Phys.}}          
\def\npa{\aaref@jnl{Nucl.~Phys.~A}}        
\def\lrr{\aaref@jnl{Living Rev.~Rel.}}                   
\def\jcap{\aaref@jnl{J.~Cosmology Astropart.~Phys.}}    
\def\rmp{\aaref@jnl{Rev.~Mod.~Phys.}}   
\def\epjc{\aaref@jnl{Eur.~Phys.~J.~C}}

\allowdisplaybreaks[1]

\begin{document}
\color{black}       
\title{Investigating Stable Quark Stars in Rastall-Rainbow Gravity and Their Compatibility with Gravitational Wave Observations}

\author{Takol Tangphati} 
\email[]{takoltang@gmail.com}
\affiliation{School of Science, Walailak University, Thasala, \\Nakhon Si Thammarat, 80160, Thailand}

\author{Dhruba Jyoti Gogoi\orcidlink{0000-0002-4776-8506}}
 \email{moloydhruba@yahoo.in}
\affiliation{Department of Physics, Moran College, Moranhat, Charaideo 785670, Assam, India.}
\affiliation{Theoretical Physics Division, Centre for Atmospheric Studies, Dibrugarh University, Dibrugarh
786004, Assam, India.}

\author{Anirudh Pradhan} 
\email[]{pradhan.anirudh@gmail.com}
\affiliation{Centre for Cosmology, Astrophysics and Space Science, GLA University, Mathura-281 406, Uttar Pradesh, India}

\author{Ayan Banerjee \orcidlink{0000-0003-3422-8233}} 
\email{ayanbanerjeemath@gmail.com}
\affiliation{Astrophysics and Cosmology Research Unit, School of Mathematics, Statistics and Computer Science, University of KwaZulu--Natal, Private Bag X54001, Durban 4000, South Africa}


\date{\today}

\begin{abstract}
We present a stable model for quark stars in Rastall-Rainbow (R-R) gravity. The structure of this configuration is obtained by utilizing an interacting quark matter equation of state.  The R-R gravity theory is developed as a combination of two distinct theories, namely, the Rastall theory and the gravity's rainbow
formalism. Depending on the model parameters ($\bar{\lambda}, \eta, \Sigma, B_{\rm eff}$), the mass-radius relations are numerically computed for modified Tolman-Oppenheimer-Volkoff (TOV) equations with proper boundary conditions. The stability of equilibrium configuration
 has been checked through the static stability criterion, adiabatic index and the sound velocity. Our calculations predict larger maximum masses for quark stars, and the obtained results are compatible with accepted masses and radii values, including constraints from GW190814 and GW170817 events in all the studied cases.

\end{abstract}

\maketitle

\section{Introduction}

Over the centuries, Einstein's theory of general relativity (GR) has stood like a pillar of modern theoretical physics \cite{Will:2014kxa}.  The success of this theory comes through the first experimental test by Sir Arthur Eddington in 1919, during a total solar eclipse.
Since then GR remains to be the most successful gravity theory for understanding the universe.  Also, GR is the simplest metric theory of gravity that passed all experimental tests at the solar system scale. Among many astonishing predictions concerning GR, compact astrophysical objects such as black holes, neutron stars, white dwarfs have turned  from purely mathematical objects to potentially real physical entities.

In 1967, the discovery of pulsars had a great impact on astronomers in general. This discovery proves the
existence of neutron stars (NSs) in the Universe and crucial to understand the nature of ultra-dense compact objects. NSs are the incredibly dense remnants of massive stars when they run out of fuel.  At this stage, the energy production stops at the core of NSs and starts rapidly collapsing, squeezing electrons and protons together to form neutrons and neutrinos. Such stars are supported by neutron-degeneracy pressure, and thus, are the most compact stars in the Universe. A NS having mass between $M \sim 1-3 M_{\odot}$ where $M_{\odot} =2\times 10^{33}$ ${\rm{g}}$ with radius between 10-15 {\rm km} \cite{Ozel:2016oaf,Steiner:2017vmg}. Thus, their central densities are extremely high and easily exceed the nuclear saturation limit i.e., $\rho \gtrsim \rho_{\text{nuc}} $ where $\rho_{\text{nuc}} = 2.8 \times 10^{14}$ $\rm g/cm^3$. It is therefore hard to deal with the matter in such an extreme situation in a laboratory conducted on Earth, and thus no comprehensive picture
has been authorized till date. 

Moreover, observed pulsars through electromagnetic (EM) signals have put a strong constraint on the equation of state (EoS) 
of dense matter in the interior of NSs. Meanwhile, the  mass-radius measurements from spectroscopic observations of thermonuclear X-ray bursts, along with recent NICER (Neutron Star Interior Composition Explorer) data have significantly placed tight constraints on the EoS further \cite{Miller:2019cac}. In particular, the detection of massive millisecond pulsar (MSP) known as PSR J0952-0607  was discovered by Bassa {\it{et al}} \cite{Bassa:2017zpe} has ruled out a large number of EOSs based on exotic degrees of freedom. For the above mentioned reasons, physicists predict the existence of more exotic states such as strange quark matter (SQM) in the
core of compact objects. This was first speculated in \cite{Itoh,Witten:1984rs,Bodmer:1971we} that compact stars could be partially or totally made of SQM. It has been suggested that SQM consists of almost equal numbers of $u$, $d$ and $s$ quarks, and a small number of electrons to attain the charge neutrality. Quarks are strongly interacting particles and may exist from a few fermis up to a large (kilometer-sized) ranging in size with the possibility of consistent self-bound quark stars (QSs). The simple model proposed for SQM is the MIT bag model \cite{Farhi:1984qu} in which the quarks are considered to be free inside a bag.  Depending on this model, the
internal structure of QSs has been explored by several authors (see, e.g., Refs. \cite{Bora:2022qwe,Bora:2023zhp,Bora:2022dnu}).

Both isotropic compact stars \cite{Bora:2022qwe,Bora:2023zhp,Bora:2022dnu,Pretel:2020oae,Zhang:2023mzb} and anisotropic compact stars \cite{Rej:2021ngp,Bhar:2020abv,Errehymy:2021cdl,Sharif:2020glz,Biswas:2020gzd,Maurya:2019sfm,Lopes:2019psm} got significant importance from different researchers. Recently, isotropic compact quark stars have been investigated in Ho\v{r}ava gravity and Einstein-\ae ther theory in which both linear and non-linear EoS, associated with the MIT bag model and colour flavour locked state have been extensively considered and investigated. The study showed how the compactness and the M-R relations are affected by the model parameter elaborately \cite{n001}. 

In Rastall's theory of gravity \cite{Rastall:1972swe}, several significant studies have been done related to compact stars \cite{Bora:2023zhp,Majeed:2023xde,ElHanafy:2022kjl,Ghosh:2021byh} and black holes \cite{Gogoi:2023ntt,Bezerra:2022srj,Meng:2020csd,Gogoi:2021cbp,Gogoi:2021dkr}. One may note that although Rastall gravity is equivalent to GR in weak field approximation or can be expressed in a GR-like form with an effective energy-momentum tensor \cite{Visser:2017gpz}, the theory deviates significantly from GR in the presence of matter or non-zero curvature \cite{Darabi:2017coc,Oliveira:2015lka}.
Rastall and Rainbow gravity theories were combined to study NSs in Ref. \cite{Mota:2019zln}. In this study, authors found that even for minute alterations of the associated model parameters, significant variations were observed. This study reveals a promising aspect of such theories and suggests compatibility with observed astrophysical data. Apart from compact stars, wormholes also have been extensively investigated in R-R gravity. In Ref. \cite{Tangphati:2023nwz}, traversable wormholes have been investigated in R-R gravity framework. This research reveals that the possibility of static and spherically symmetric wormholes emerging in a zero-tidal-force setting is not attainable for specific combinations of free parameters and equations of state. The authors, focusing on the subset of viable solutions, systematically evaluate their stability using adiabatic sound velocity analysis and assess their adherence to the Weak Energy Condition (WEC). In essence, this investigation sheds light on how the interaction between Rastall parameters and Rainbow functions could mitigate violations of energy conditions in these modified gravity scenarios. In another recent study, non-commutative effects on wormholes in R-R gravity have been investigated \cite{Pradhan:2023vhn}. Here noncommutativity was implemented through the adoption of two different distributions of energy density (Gaussian and Lorentzian) in the Morris and Thorne metric. In this case, particularly noteworthy is the observation that, within specific parameter ranges, it becomes possible to mitigate the violation of the WEC at the throat and in the vicinity of the wormholes in R-R gravity framework.

Motivated by these studies, here we focus on the isotropic case of quark stars in R-R gravity, which has been studied extensively in different gravity theories due to their interesting results as well as their comparative mathematical simplicity. Isotropic stars, in this context, are characterized by a uniform distribution of key attributes, with pressure solely dependent on density. This simplification allows for a more straightforward mathematical treatment, making isotropic quark stars a valuable subject of analysis.

This investigation aims to provide a comprehensive understanding of the structural aspects of these stars by focusing on the isotropic case, giving insight into their behaviour and characteristics in astrophysical situations in R-R gravity. It is focused on contributing to our understanding of the properties of isotropic quark stars in this gravity theory, furthering our understanding of the Universe's phenomena. This investigation also delves into the constraints imposed on compact quark stars by recent gravitational wave observations, focusing particularly on the significance of GW190814 \cite{LIGOScientific:2020zkf} and GW170817 \cite{LIGOScientific:2018cki}. The breakthrough moment occurred on August 17, 2017, when the LIGO and Virgo observatories directly detected gravitational waves stemming from the coalescence of a binary neutron star system \cite{LIGOScientific:2018cki}. The subsequent observation of GW190814 during the third observing run in 2019 added another layer of insight, boasting a remarkable signal-to-noise ratio of 25 in the three-detector network. This event, characterized by an unprecedented unequal mass ratio in gravitational wave measurements, introduces the secondary component as potentially the lightest black hole or the heaviest neutron star ever identified in a double compact-object system \cite{LIGOScientific:2020zkf}. The findings from these gravitational wave signals, associated with potential compact stars, hold a pivotal role in shaping and refining theoretical models of various compact stars, offering a nuanced understanding of extreme conditions.

The structure of our work unfolds as follows: In Section \ref{sec2}, we provide a concise overview of R-R gravity theory, accompanied by an exploration of the hydrostatic equilibrium equations governing stellar systems within the framework of R-R gravity. Moving on to Section \ref{sec3}, we delve into the EoS for interacting quark matter. Section \ref{sec4} is dedicated to the presentation and analysis of numerical results, emphasizing the influence of model parameters on stellar structures. Within this context, we scrutinize stability conditions in Section \ref{sec5}. To conclude, our findings and insights are encapsulated in Section \ref{sec6}. Throughout the study, we employ the geometric unit system, yet we present our results in physical units to facilitate meaningful comparisons.

\section{Field equations of Rastall-Rainbow gravity}\label{sec2}

Let us start by discussing the unification of Rastall and rainbow theories. The Rastall-Rainbow (R-R) gravity model \cite{Mota:2019zln} is another example of modified gravity theory, which consists of two modified theories, namely the Rastall theory \cite{Rastall:1972swe} and the Rainbow theory \cite{Magueijo:2002xx}.

\subsection{Rainbow theory}
Based on the generalization of doubly special relativity to the curved spacetime, Magueijo and Smolin \cite{Magueijo:2002xx}
arrive a new gravity theory called gravity's rainbow. In this framework, the geometry of the spacetime depends on the energy of the test particles in it.   Hereby, particles with different energies distort spacetime differently that arise a modification of energy-momentum dispersion relation, which is 
\begin{equation}
E^{2} \Xi(x)^{2} - p^{2}\Sigma(x)^{2} = m^{2},
\label{eq1}
\end{equation}
It is notable that the expression $x = E/E_{p}$ represents the dimensionless ratio between an energy $E$ of the probe particle
of mass $m$ and $E_{p}$ is the Planck energy. Here, the two functions $\Xi(x)$ and $\Sigma(x)$ are known as rainbow functions and play a vital role in the Rainbow gravity framework. This modified form of the relativistic dispersion relation is significant in the ultraviolet limit and in low energy levels the rainbow functions $ \Xi(x)$ and $\Sigma(x)$ are chosen so that $x = E/E_{p} \rightarrow 0 $, and  these functions go to unity, i.e., 
\begin{equation}
    \lim_{x\rightarrow 0} \Xi(x)=1, \quad \lim_{x \rightarrow 0} \Sigma(x)=1.
    \label{eq2}
\end{equation}
with restoring the standard dispersion relation.

Following \cite{Magueijo:2002xx}, the energy-dependent metric in the following form
\begin{equation}
g^{\mu\nu}(x)=\eta^{ab} e_{a}^{\mu}(x)\otimes e_{b}^{\nu}(x),
\label{eq3}
\end{equation}
where $\eta^{ab}$ is the Minkowski tensor with the energy-dependent vierbein fields $e_{a}^{\mu}(x)$ are related through the energy independent frame fields by the following expressions:
\begin{equation}
e_{0}^{\mu}(x)=\frac{1}{\Xi(x)} \widetilde{e}_{0}^{\mu}, \quad e_{k}^{\mu}(x)=\frac{1}{\Sigma(x)} \widetilde{e}_{k}^{\mu}.
\label{eq4}
\end{equation}
the index runs from $k = (1, 2, 3)$ represents the spatial coordinates. With this methodical proposal, one can modify the Einstein's field equations to the energy-dependent Einstein field equations, and leads to a change in the static spherically symmetric metric  to energy dependent metric 
by using Eq. (\ref{eq3}) and considering the quantities  $\widetilde{e_{i}}$, 
\begin{eqnarray}\label{metric_RR}
    ds^2 = - \frac{B(r)}{\Xi^2(x)} dt^2 + \frac{A(r)}{\Sigma^2(x)}dr^2 + \frac{r^{2}}{\Sigma^{2}(x)} d\Omega^2_2,  
\end{eqnarray}
where $d\Omega^2_2=(d\theta^{2}+\sin{\theta}^{2}d\phi^{2})$ is the standard metric on the unit 2-sphere with $A(r)$ and $B(r)$ are the metric potentials depend on the radial coordinate $r$. In addition, the standard spherical coordinate $r$, $t$, $\theta$ and $\phi$ are independent of the energy of the probe particles. In the next phase we will investigate the effect of the energy dependence in the context of Rastall gravity.

\subsection{Rastall theory}

Rastall gravity theory is a simple generalization of GR that has been proposed by Rastall \cite{Rastall:1972swe} in 1972. The basic argument of this theory is the violation of the usual conservation law in a curved spacetime, which differs from the standard GR i.e., $T^{\mu \nu}_{\quad;\mu}=0$. Interestingly, the left side of the usual Einstein's field equations holds
the Bianchi identity i.e., $G^{\mu \nu}_{\quad;\mu}=0$. Rastall's theory is based on the following assumption that the divergence of ($T_{\mu\nu}$) is proportional to the gradient of the curvature scalar ($R$). In this framework, the proposed modified conservation law given by Rastall is expressed as \cite{Rastall:1972swe}:  
\begin{equation}
    T_{\: \: \mu;\nu}^{\nu}=\Bar{\eta}R_{,\mu},
    \label{conserve_eqn}
\end{equation}
where $\Bar{\lambda}$ is an undetermined constant. The Eq. (\ref{conserve_eqn}) can be written as
\begin{equation}
    \left(T_{\: \: \mu}^{\nu}-\Bar{\eta}\delta_{\: \:\mu}^{\nu}R\right)_{;\: \nu} =0.
    \label{eq13}
\end{equation}
In this way, there exists a non-minimal coupling between matter and geometry through the following field equations 
\begin{equation}
  R^{\nu}_{\: \: \mu}-\frac{1}{2}\delta_{\: \:\mu}^{\nu}R= 8\pi G\left(T_{\: \: \mu}^{\nu}-\Bar{\eta}\delta_{\: \:\mu}^{\nu}R\right).  
  \label{eq14}
\end{equation}
Finally, we can rewrite the above equation in more convenient form where the energy-moment tensor stays on the right side, i.e.,
\begin{equation}
    R^{\nu}_{\: \: \mu}-\frac{\eta}{2}\delta_{\: \:\mu}^{\nu}R= 8\pi G T_{\: \: \mu}^{\nu},
  \label{eq9} 
\end{equation}
where $\bar{\eta} = \frac{1 - \eta}{16 \pi G}$. If we take the $\eta =1$ limit, we recover the standard equation of motion of GR. The parameter $\eta $ is called the Rastall parameter and leads to the generalization of the Einstein's equation. 

\subsection{Rastall-Rainbow theory}

In \cite{Mota:2019zln}, the authors pointed out that one can construct another modified gravity theory by combining both theories discussed above. This theory is called the Rastall-Rainbow (R-R ) gravity theory, and the field equations of this model can be incorporated into Eq. (\ref{eq9}) by considering an energy-dependent metric and gravitational constant $G(x)$.  Following \cite{Mota:2019zln}, the equation of motion for R-R gravity is 
\begin{equation}
R_{\mu}{}^{\nu}(x) - \frac{\eta}{2}\delta_{\mu}{}^{\nu}(x)R(x) = k(x) T_{\mu}{}^{\nu}(x)  
 \label{eq16},
\end{equation}
where $k(x) = 8 \pi G(x)$ and $G(x)$ represents the energy-dependent gravitational constant.  The effects of such modifications can lead to an interplay between gravity, quantum theory, and the underlying structure of spacetime. Here, we assume $G(x)= c(x)=1$  throughout the paper.

To make the above Eq. (\ref{eq16}) in a more compactified  form, we add and subtract the term $(1/2) g_{\mu\nu}R$ to the left side, which turns out the usual Einstein equation with an effective energy-moment tensor in the right-hand side, as  
 \begin{equation}
     R_{\mu\nu}-\frac{1}{2}g_{\mu\nu}R=8\pi \tau_{\mu \nu},
     \label{eq13b}
 \end{equation}
 where we define
 \begin{equation}
     \tau_{\mu\nu}=T_{\mu\nu}-\frac{(1-\eta)}{2(1-\eta)}g_{\mu\nu}T.
      \label{eq13c}
 \end{equation} 
In the following discussion, it will be interesting to the possibility of addressing some problems concerning the internal structure of a QS in R-R gravity. For, this purpose 
we consider the perfect fluid form of the EMT given by
\begin{eqnarray}
 T_{\mu \nu} = (\rho + p)u_\mu u_\nu+ p g_{\mu\nu}, \label{EnMoIso}
\end{eqnarray}
where $\rho(r)$ is the  energy density, $p(r)$ is the pressure of the fluid, and $u_{\mu}$ is the 4-velocity satisfying the conditions
\begin{equation}
    u_{\mu}=\left( \frac{\Xi(x)}{\sqrt{B(r)}},0,0,0\right)
    \label{eq13}.
\end{equation}
Using the metric given in Eq. (\ref{metric_RR}) with the EMT (\ref{EnMoIso}), we reach the following $(tt)$ and $(rr)$ components,
\begin{eqnarray}
    M'(r) &=& 4 \pi r^2 \Tilde{\rho}, \label{master1} \\
    \frac{1}{r}\left( 1 - \frac{2 G M(r)}{r} \right) \frac{B'(r)}{B(r)} - \frac{2  M(r)}{r^3} &=& 8 \pi \Tilde{p}, \label{master2}
\end{eqnarray}
Here, we write the metric potential $A(r)$ in term of the mass function $M(r)$ given by  $A(r)^{-1} = 1 - \frac{2  M(r)}{r}$, and $ \Bar{\rho}$ and $\Bar{p}$ represent the effective energy density and  pressure in the form
\begin{eqnarray}
   \Tilde{\rho } &= \frac{1}{ 2\Sigma^2(x) ( 2 \eta - 1)} \left( (3\eta - 1) \rho + 3 (\eta - 1) p \right), \\
    \Tilde{p} &= \frac{1}{ 2\Sigma^2(x) ( 2 \eta - 1)} \left( (\eta + 1) p + (\eta - 1) \rho \right).
\end{eqnarray}
This effective density and pressure are depending on the new parameters $\eta$ and $\Sigma$. The case of $\eta = 1$ and $\Sigma=1$ corresponds to the usual definition of the GR. We consider the radial component of Eq.~(\ref{conserve_eqn}) and apply Eq.~(\ref{master2}) to eliminate the function $B(r)$, which gives \cite{Mota:2019zln}
\begin{eqnarray}\label{master3}
    \frac{d \Bar{p} }{dr} = - \frac{M + 4 \pi \Bar{p} r^3}{r^2 \left( 1 - \frac{2  M}{r} \right)} \left( \bar{\rho} + \bar{p} \right).
\end{eqnarray}
This represents the stellar hydrostatic equilibrium equation within the framework of R-R gravity. So, the final three differential equations needed to be solved are (\ref{master1}), (\ref{master2}) and (\ref{master3}) with an EoS for the star matter $\bar{p}= \bar{p}(\bar{\rho})$. In the next section, we present the structural equation that describes the interior of QS configurations. 

 \section{ Interacting quark matter EoS } \label{sec3}

Here, we start by considering an interacting quark matter EoS that includes interquark effects such as perturbative QCD (pQCD) corrections and color superconductivity \cite{Zhang:2020jmb}. It is quite remarkable that depending only on a single parameter one can rescale the EoS into a dimensionless form which characterizes the size of strong interaction effects. The main motivation of this article is to utilize the interacting quark matter EoS unifying all macroscopic properties of QSs.

Within this framework, we start by writing the relation between the energy density  $(\rho)$ and the pressure $(p)$ as follows \cite{Zhang:2020jmb,Zhang:2021fla}: 
{\fontsize{9.7}{12}
\begin{align}
p=\frac{1}{3}(\rho-4B_{\rm eff})+ \frac{4\lambda^2}{9\pi^2}\left(-1+{\rm sgn}(\lambda)\sqrt{1+3\pi^2 \frac{(\rho-B_{\rm eff})}{\lambda^2}}\right),
\label{eos_tot}
\end{align}}
where $B_{\rm eff}$ stands for the effective bag constant that accounts for the nonperturbative contribution from the QCD vacuum and
\begin{eqnarray}
\lambda=\frac{\xi_{2a} \Delta^2-\xi_{2b} m_s^2}{\sqrt{\xi_4 a_4}}.
\label{lam}
\end{eqnarray}
In the above expression, the notations $\Delta$ and $m_s$ represent the gap parameter and the strange quark mass, respectively. The coefficient $a_4$ is parameterization of QCD corrections from one-gluon exchange for gluon interaction to $O{(\alpha_s^2)}$,  and varies from small values to $a_4=1$. The sign of $\lambda$ is represented by ${\rm sgn}(\lambda)$, and positive as long as $\Delta^2/m_s^2>\xi_{2b}/\xi_{2a}$. The constant coefficients in $\lambda$ are
\begin{figure}
    \centering
    \includegraphics[width = 8.4 cm]{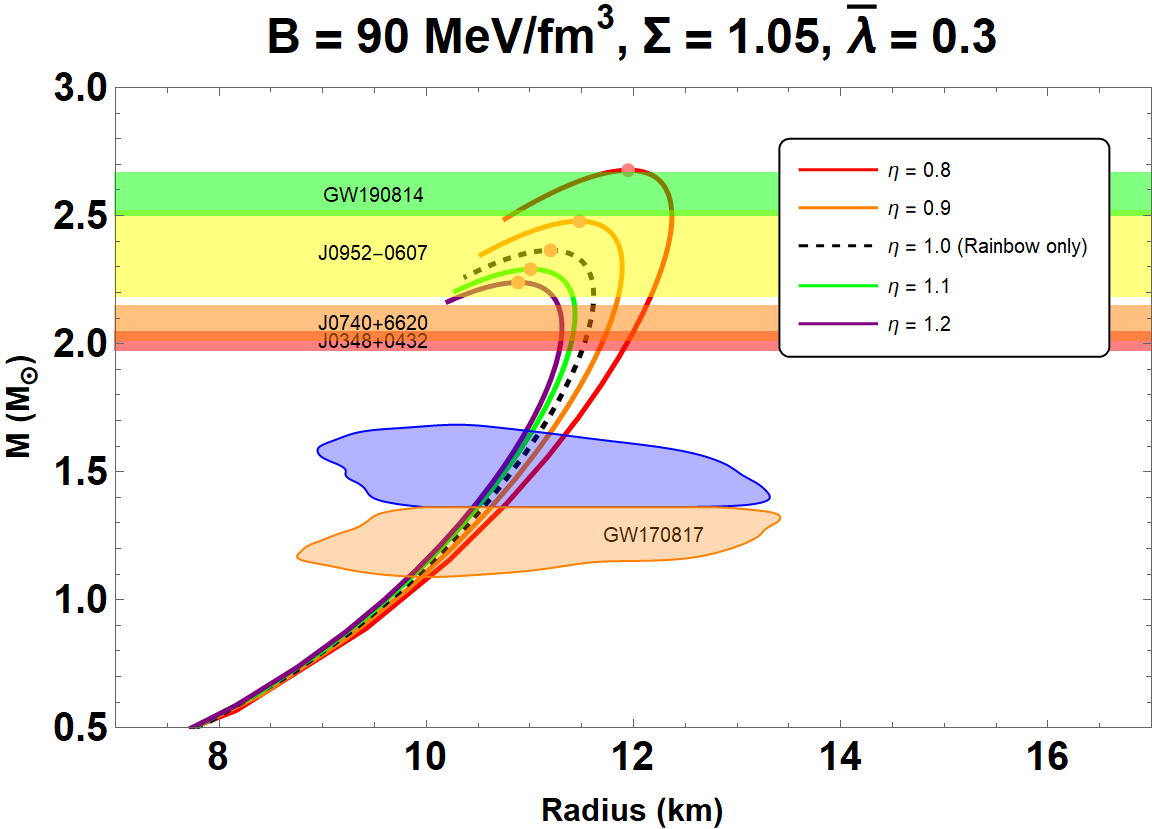}
    \includegraphics[width = 8.4 cm]{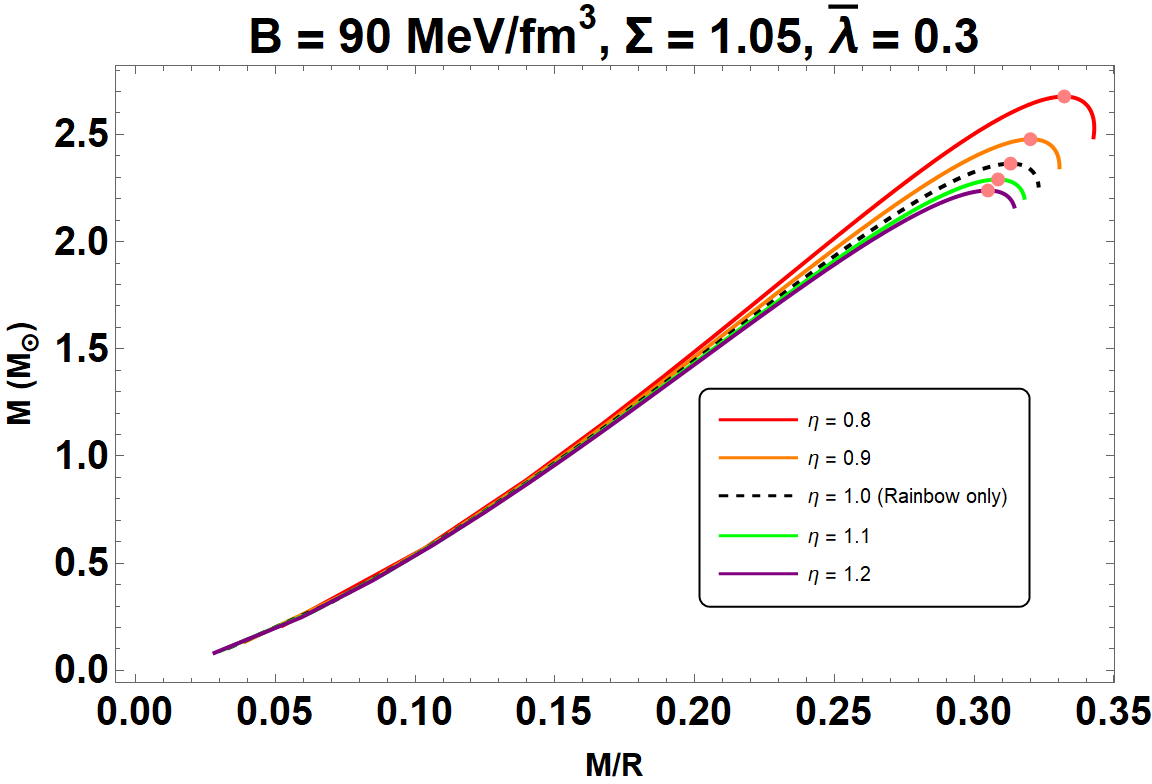}
    \caption{The profiles of mass-radius and mass-compactness relations of QSs in the R-R gravity theory with the given EoS (\ref{eos_p}). The parameters used in the model described by $B_{\rm eff} = 90$ MeV/fm$^3$, $\Sigma = 1.05$, $\bar{\lambda} = 0.3$, and $\eta \in$ [0.8, 1.2]. To compare with observational constraints, we consider massive pulsars data such as PSR J0952-0607 \cite{Romani:2022jhd}, PSR J0740+6620  \cite{Fonseca:2021wxt}  and PSR J0348+0432  \cite{Antoniadis:2013pzd} with different colour bars. Furthermore, we include the constraint from GW190814 \cite{LIGOScientific:2020zkf} and GW170817 event \cite{LIGOScientific:2018cki}, also.}
    \label{fig_profiles_vary_eta}
\end{figure}
\begin{align}
(\xi_4,\xi_{2a}, \xi_{2b}) = \left\{ \begin{array} {ll}
(( \left(\frac{1}{3}\right)^{\frac{4}{3}}+ \left(\frac{2}{3}\right)^{\frac{4}{3}})^{-3},1,0) & \textrm{2SC phase}\\
(3,1,3/4) & \textrm{2SC+s phase}\\
(3,3,3/4)&   \textrm{CFL phase}
\end{array}
\nonumber
\right.
\end{align}
that characterizing the possible phases of color superconductivity. As of Ref. \cite{Zhang:2020jmb}, we now introduce the dimensionless  rescaling:
\begin{align}
\bar{\rho}=\frac{\rho}{4\,B_{\rm eff}}, \,\, \bar{p}=\frac{p}{4\,B_{\rm eff}},  \,\,
\label{rescaling_prho}
\end{align}
and 
\begin{align}
 \bar{\lambda}=\frac{\lambda^2}{4B_{\rm eff}}= \frac{(\xi_{2a} \Delta^2-\xi_{2b} m_s^2)^2}{4\,B_{\rm eff}\xi_4 a_4}.
 \label{rescaling_lam}
\end{align}
After introducing the rescaling (\ref{rescaling_prho}) and (\ref{rescaling_lam}), we finally have the dimensionless form of Eq.~(\ref{eos_tot}), which is 
\begin{align}
\bar{p}=\frac{1}{3}(\bar{\rho}-1)+ \frac{4}{9\pi^2}\bar{\lambda} \left(-1+{\rm sgn}(\lambda)\sqrt{1+\frac{3\pi^2}{\bar{\lambda}} {(\bar{\rho}-\frac{1}{4})}}\right).
\label{eos_p}
\end{align}
It is easy to see that when $\bar{\lambda} \to 0$, the 
Eq. (\ref{eos_p}) become $\bar{p}=\frac{1}{3}(\bar{\rho}-1)$, which represent the conventional noninteracting rescaled quark matter EoS. In fact, when we consider extremely large positive values of  $\bar{\lambda}$, the Eq. (\ref{eos_p}) has the special form 
\begin{align}
\bar{p}\vert_{\bar{\lambda}\to \infty}=\bar{\rho}-\frac{1}{2}. 
\label{eos_infty1}
\end{align}
The Eq.~(\ref{eos_infty1}) is equivalent to $p={\rho}-2B_{\rm eff}$ after scaling back by using the Eq.~(\ref{rescaling_prho}). However, the Eq. (\ref{eos_p}) does not have a finite form for a negative value of $\lambda$, as 
$\bar{\lambda} \to \infty$. As explained in Ref \cite{Zhang:2020jmb,Zhang:2021fla}, the positive increasing values of $\bar{\lambda}$ give a stiffer EoS and sufficiently high masses \cite{Zhang:2020jmb,Zhang:2021fla} for QSs. In the subsequent sections, we solve the field equations numerically for the given EoS. 


\section{Numerical results and discussion} \label{sec4}

\begin{figure}[h]
    \centering
    \includegraphics[width = 8.4 cm]{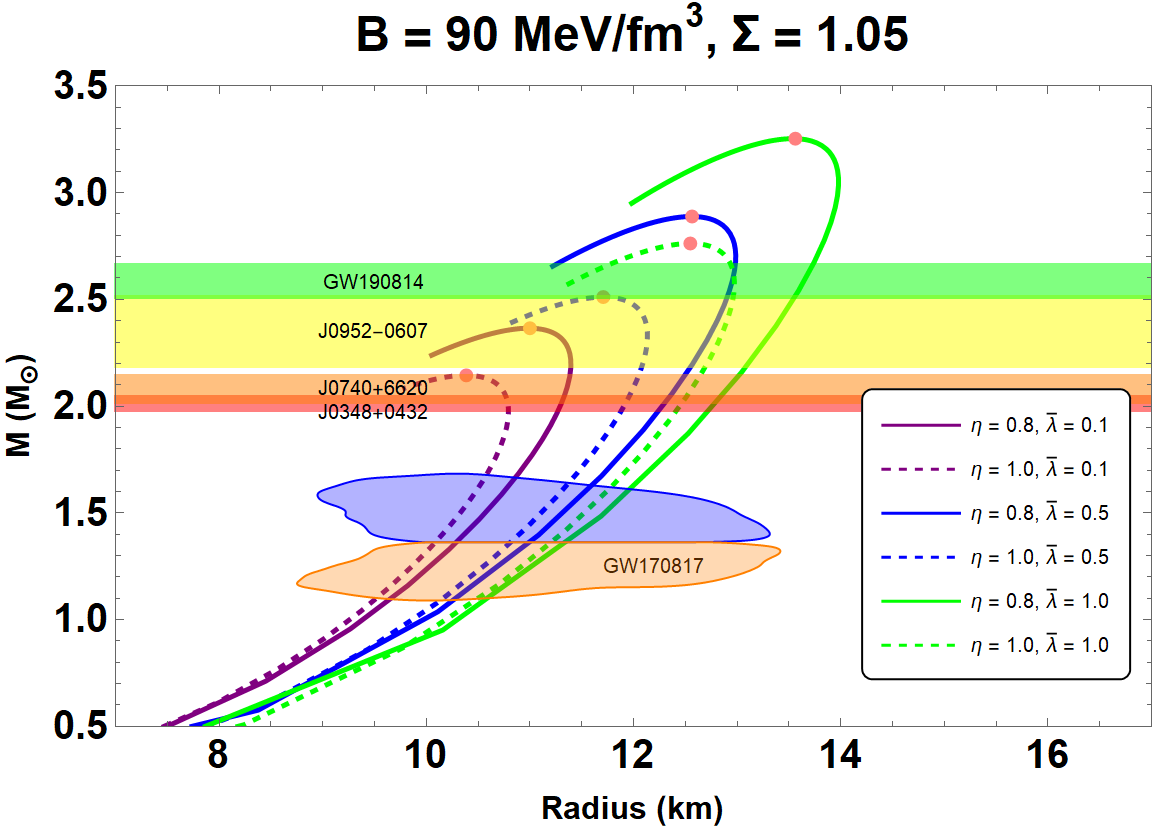}
    \includegraphics[width = 8.4 cm]{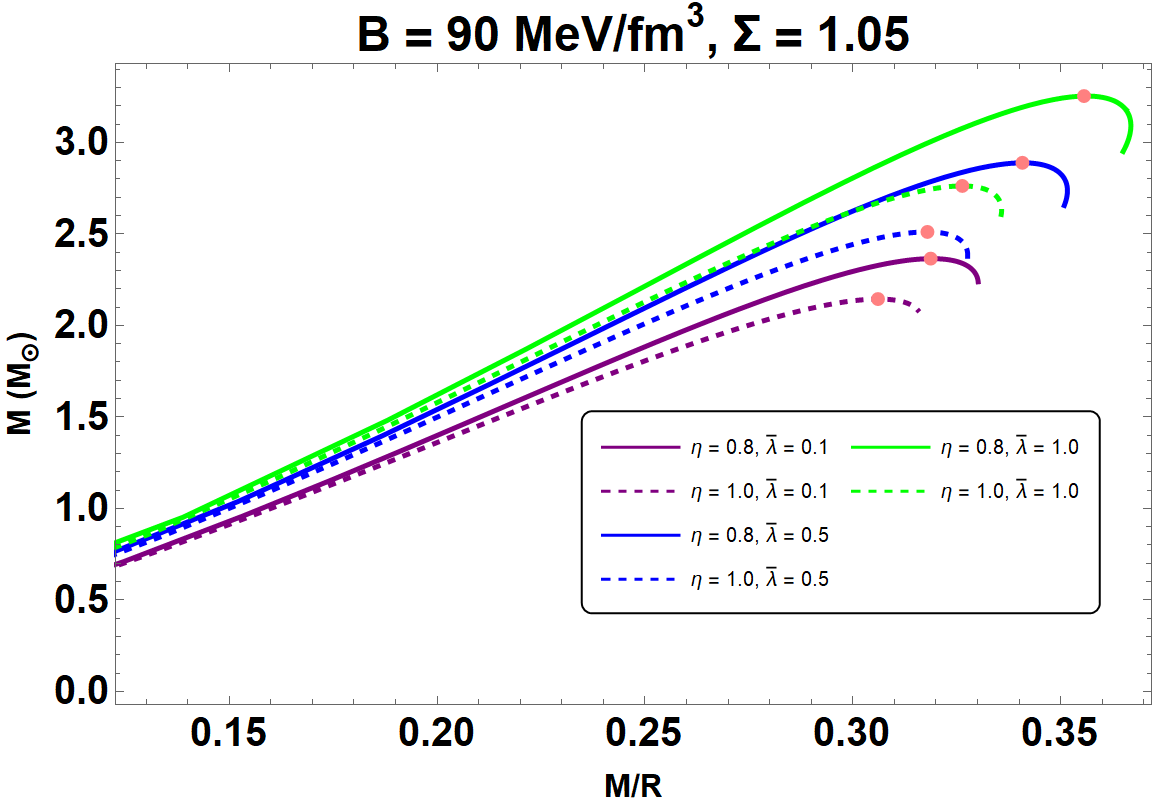}
    \caption{The profiles of mass-radius and mass-compactness relations of QSs in the R-R gravity theory with the given EoS (\ref{eos_p}). The parameters used in the model described by $B_{\rm eff} = 90$ MeV/fm$^3$, $\Sigma = 1.05$, $\bar{\lambda} \in $ [0.1, 1.0], and $\eta = \{0.8, 1.0\}$, respectively. Used constraints are the same as of Fig. \ref{fig_profiles_vary_eta}. }
    \label{fig_profiles_vary_lambda}
\end{figure}

\begin{figure}[h]
    \centering
    \includegraphics[width = 8.4 cm]{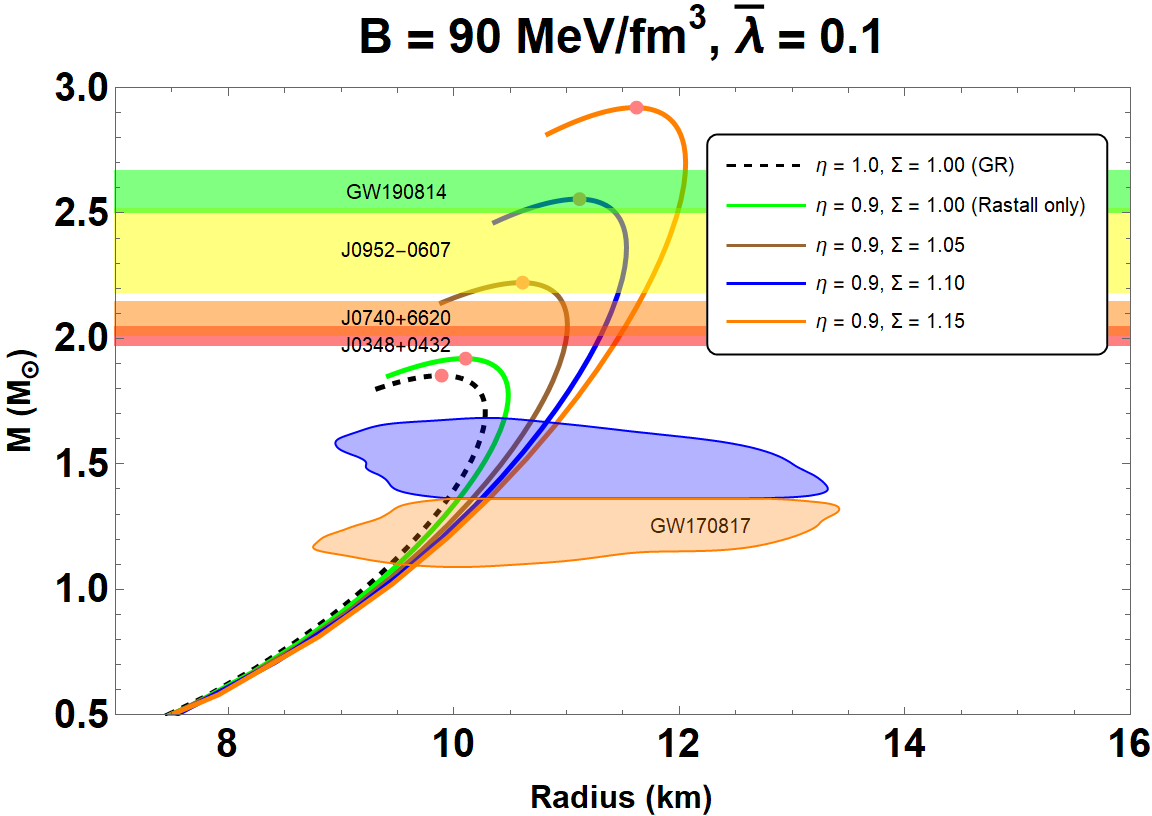}
    \includegraphics[width = 8.4 cm]{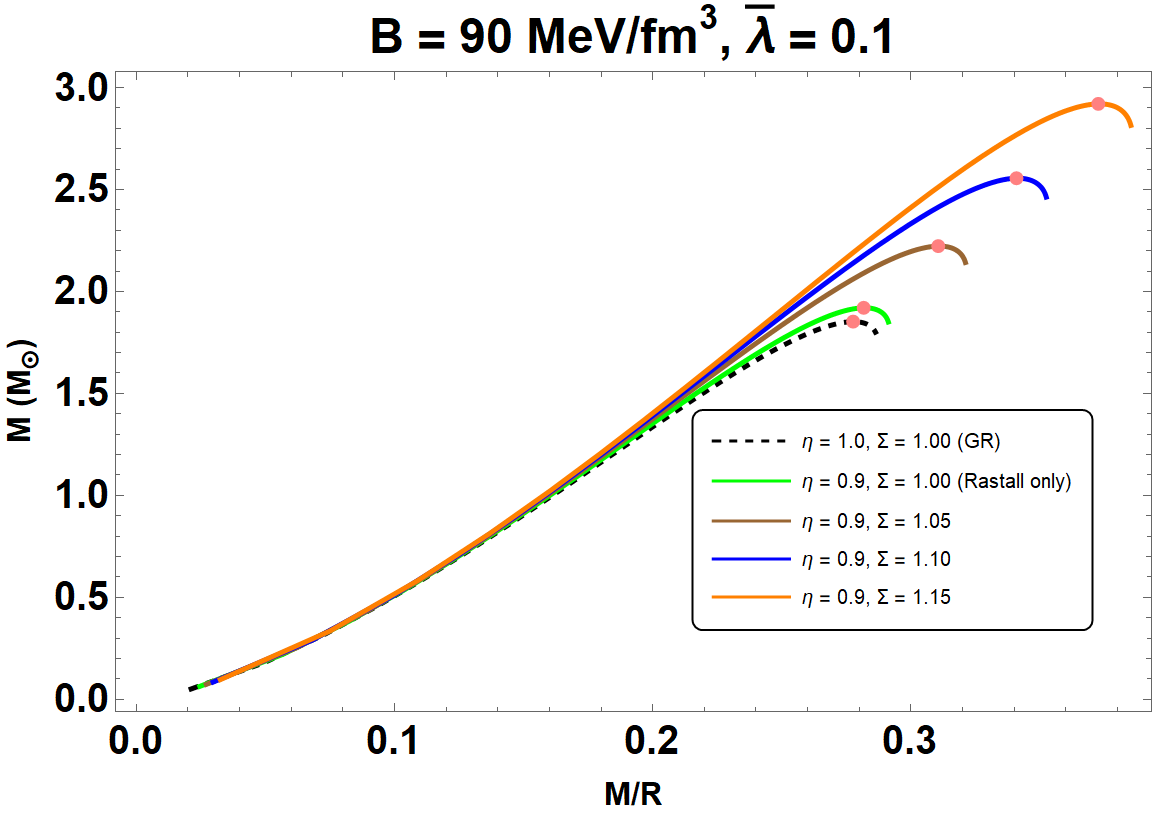}
    \caption{The profiles of mass-radius and mass-compactness relations of QSs in the R-R gravity theory with the given EoS (\ref{eos_p}). The set of  parameter are: $B_{\rm eff} = 90$ MeV/fm$^3$, $\Sigma \in $ [1.00, 1.15], $\bar{\lambda} = 0.1 $ and $\eta = 0.9$, respectively. Colors bands indicate observational constrains based on the current astrophysical data, same as of Fig. \ref{fig_profiles_vary_eta}.}
    \label{fig_profiles_vary_sigma}
\end{figure}

\begin{figure}[h]
    \centering
    \includegraphics[width = 8.4 cm]{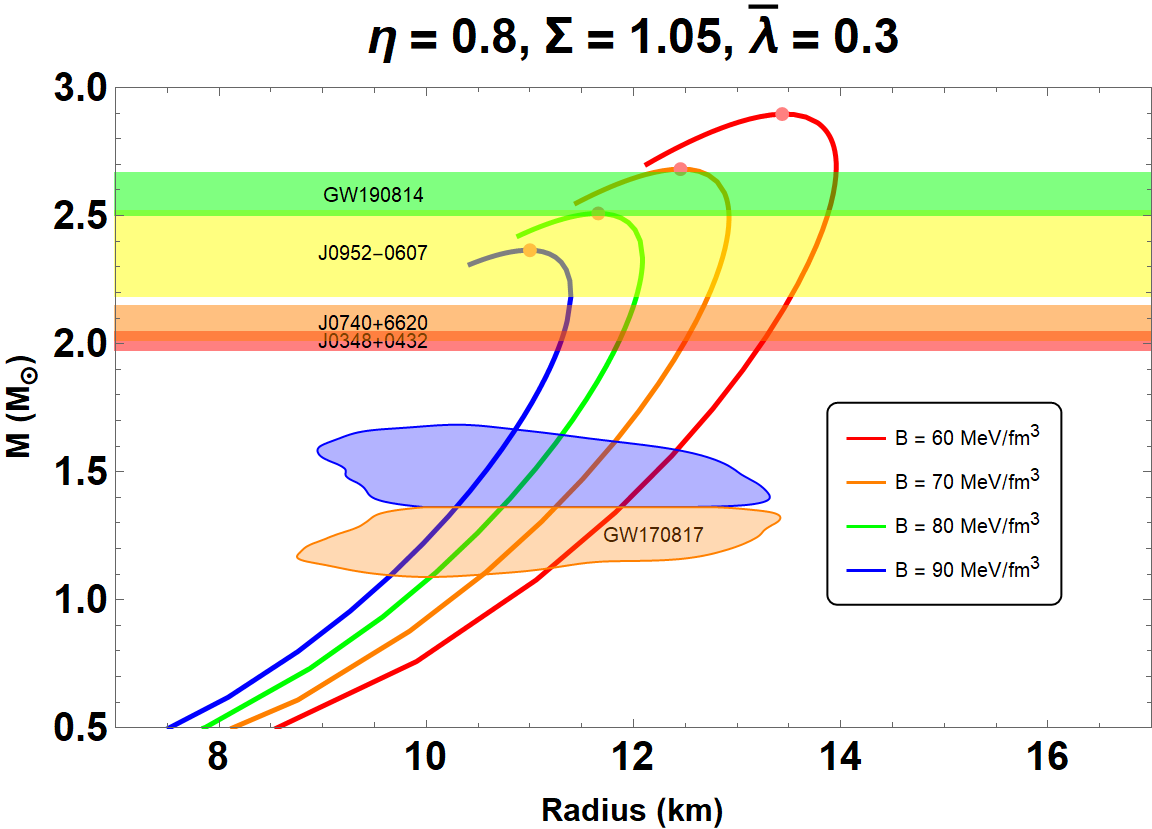}
    \includegraphics[width = 8.4 cm]{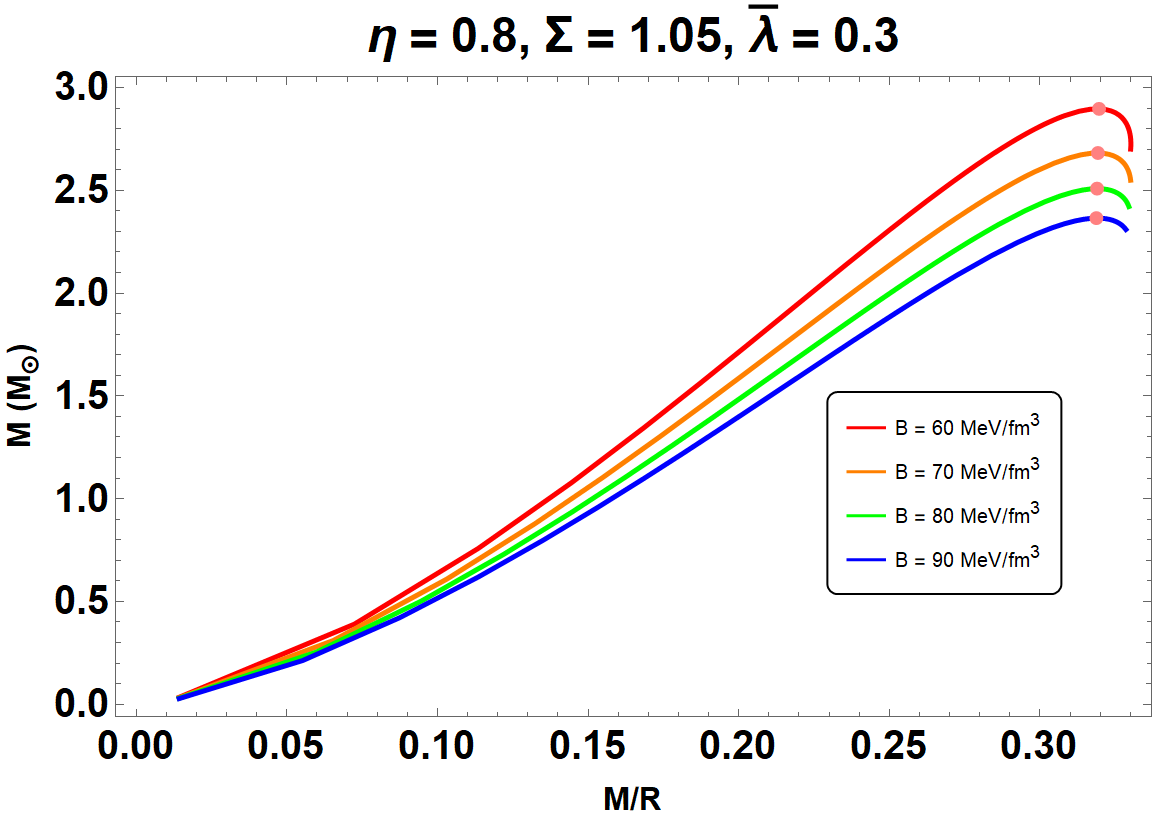}
    \caption{The profiles of mass-radius and mass-compactness relations of QSs in the R-R gravity theory with the given EoS (\ref{eos_p}). Here, we vary the effective bag constant  $B_{\rm eff} \in [60,90]$ MeV/fm$^3$ with the other parameters are $\Sigma = 1.05$, $\bar{\lambda} = 0.1 $ and $\eta = 0.8$, respectively. To check the consistency of our model, we consider observational constraints based on the current astrophysical data, same as of Fig. \ref{fig_profiles_vary_eta}. }
    \label{fig_profiles_vary_B}
\end{figure}

\begin{figure*}
	\centering
	\includegraphics[width = 8 cm]{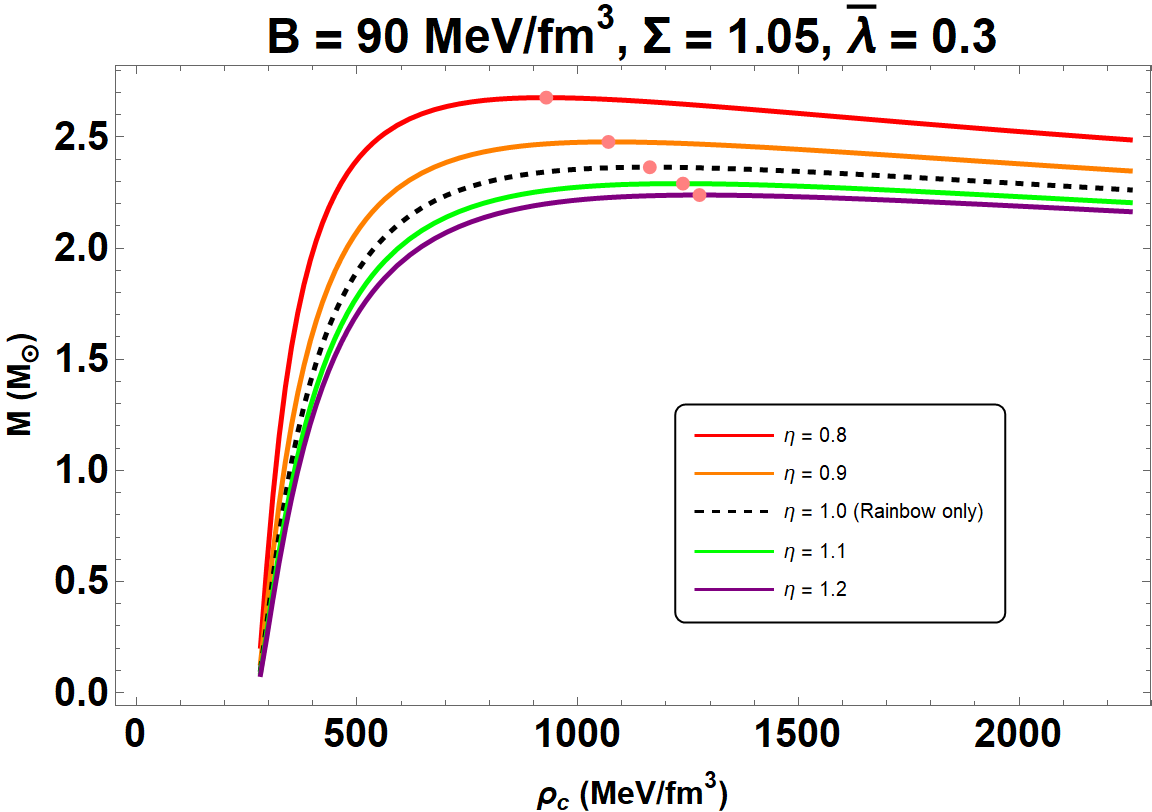}
	\includegraphics[width = 8 cm]{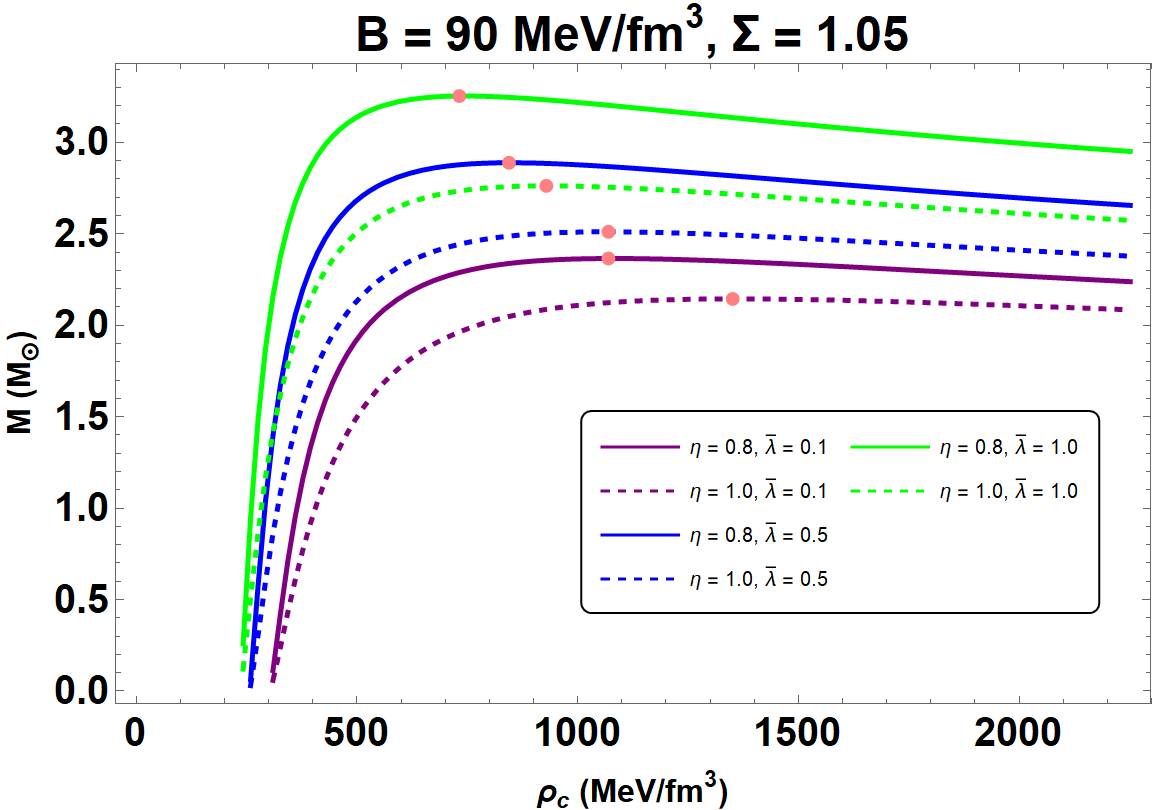}
	\includegraphics[width = 8 cm]{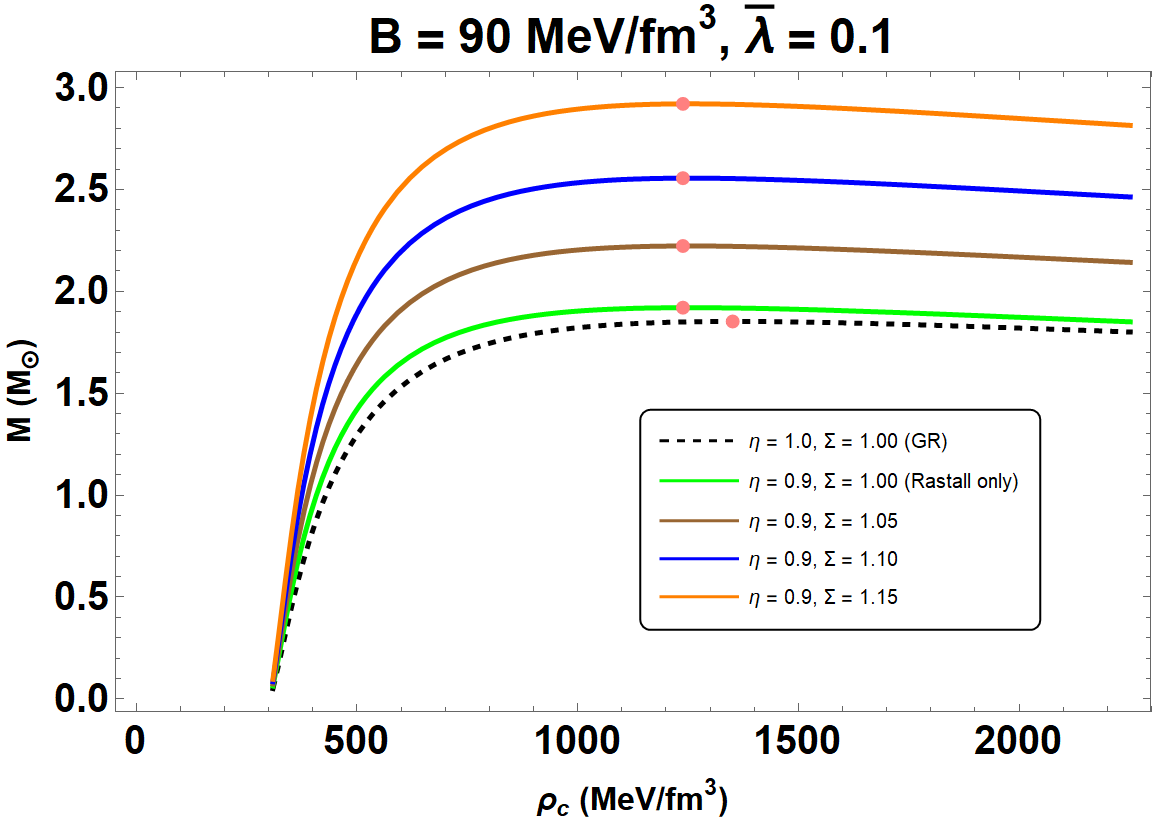}
    \includegraphics[width = 8 cm]{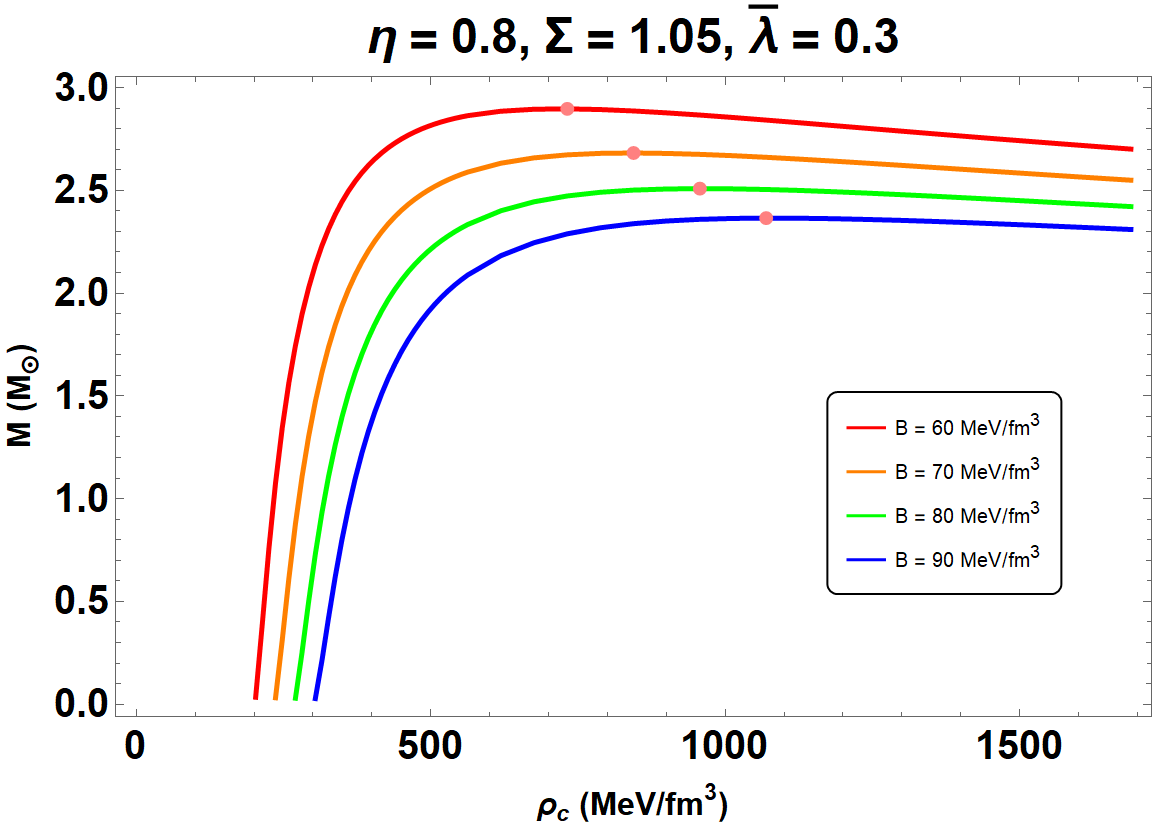}
	\caption{The profiles of mass-central energy density $(M-\rho_c)$ relations of QSs in the R-R gravity theory for the given EoS (\ref{eos_p}). Used parameters are the same as of Fig.~\ref{fig_profiles_vary_eta} to Fig.~\ref{fig_profiles_vary_B}.}
	\label{fig4}
\end{figure*}

In the following section, we solve the governing modified TOV equations (\ref{master1}) and (\ref{master3}) numerically to obtain the mass-radius relationship for QSs and explore their internal physical properties. Solutions of these equations must be sought which satisfy the boundary conditions to maintain the regularity at the stellar origin 
\begin{eqnarray}
    M(r = 0) = 0, \text{ and } \rho(r = 0) = \rho_c. \label{inner_BC}
\end{eqnarray}
where $\rho_c$ is the central energy density and varying $\rho_c$ will give different masses and radii of the star. Then we start numerical integration from the center and go up to the radial coordinate where pressure vanishes i.e., $p(r = R) = 0$. This point is defined as the star radius, $R$.

Additionally, boundary conditions are required to match the interior geometry to a spherically symmetric vacuum solution, which is defined by
\begin{equation}
B(R) = 1-\frac{2\mathcal{M}}{R},  
\end{equation}
with $\mathcal{M} = M(r=R)$ being the total mass of the star.

\subsection{Profiles for variation of Rastall free parameter $\eta$}

\begin{table}
\caption{\label{table1} 
Summary of the resulting properties of the isotropic QSs for 
$B_{\rm eff} = 90$ MeV/fm$^3$, $\Bar{\lambda} = 0.3$, $\Sigma = 1.05$, and $\eta \in [0.8, 1.2]$. }
\begin{ruledtabular}
\begin{tabular}{ccccc}
$\eta$  & $M$ [$M_\odot$]  &  $R_{M}$  [\rm{km}]  & $\rho_c$ [MeV/fm$^3$] & $M/R$  \\
\colrule
0.8  &  2.68  &  11.19  & 929 &  0.332  \\
0.9  &  2.48  &  11.48  & 1,070 &  0.320  \\
1.0  &  2.36  &  11.20  & 1,163 &  0.313  \\
1.1  &  2.29  &  11.01  & 1,238 &  0.309  \\
1.2  &  2.24  &  10.89  & 1,275 &  0.305
\end{tabular}
\end{ruledtabular}
\end{table}

In this analysis, we will explore the effect of the parameters ($\Sigma, \eta, \Bar{\lambda}$) on static QSs. We solve the structural equations for QSs using the EoS (\ref{eos_p}) and display mass-radius curves $(M-R)$, the mass-central density relations ($M-\rho_c$) and the compactness relations $(M-M/R)$ in Fig.~\ref{fig_profiles_vary_eta}. For the numerical computations, we have chosen to work with $B_{\rm eff} = 90$ MeV/fm$^3$, $\Bar{\lambda} = 0.3$, $\Sigma = 1.05$, and $\eta \in [0.8, 1.2]$. The top panel shows the results for maximum masses and their corresponding radii increase as the value of $\eta$ decreases. Table \ref{table1} summarizes the maximum mass corresponding to its radius, central energy density and compactness of QS taking into account five different values of $\eta$ from which we can quantify how QSs are affected by the variation of $\eta$. In all estimates in Table \ref{table1}, we see that increasing values of $\eta$ lead to high central energy density. The focal point of the upper plot in Fig.~\ref{fig_profiles_vary_eta} is demonstrating constraints coming from more recent observational data: PSR J0952-0607 with mass $M = 2.35 \pm 0.17 M_{\odot}$ (Yellow) \cite{Romani:2022jhd}, PSR J0740+6620 with the pulsar mass $ M = 2.08 \pm 0.07$ $M_{\odot}$ (Orange) \cite{Fonseca:2021wxt} and PSR J0348+0432 with the mass of $ M = 2.01 \pm 0.04 M_{\odot}$ (Red) \cite{Antoniadis:2013pzd}. Furthermore, we have included the constraint from the GW190814's secondary component with a mass of $2.59^{+0.08}_{-0.09} M_{\odot}$ (Green) \cite{LIGOScientific:2020zkf} and GW170817 event (M1 in blue shaded area and M2 orange shaded area) \cite{LIGOScientific:2018cki}. In this discussion we obtain the highest maximum mass for the considered parameter set is $M_{\rm{max}}$=2.68 $M_{\odot}$ with radius $R=11.19$ km for $\eta = 0.8$. In the lower panel of Fig.~\ref{fig_profiles_vary_eta}, we plot $(M-M/R)$ diagram  for the given EoS (\ref{eos_p}). According to this plot, we see larger values of $\eta$ decrease the maximum mass and maximum compactness of stars. Furthermore, in  Table \ref{table1}, we tabulated the data for the maximum compactness which lies within the range of  $0.305 < M/R < 0.332$.  The dashed black ($\eta = 1.0$) line represents the effect of Rainbow gravity theory only, and its maximum compactness is $M/R= 0.313$.


\subsection{Profiles for variation of the interaction strength parameter $\bar{\lambda}$}

In the second scenario, we describe the effect of the interaction strength parameter $\bar{\lambda}$ on the $(M-R)$ and  $(M-M/R)$ relations. For the numerical computations, we have chosen to work with $B_{\rm eff} = 90$ MeV/fm$^3$, $\Sigma = 1.05$, $\bar{\lambda} \in $ [0.1, 1.0] and $\eta = \{0.8, 1.0\}$, respectively. We have studied the dependence of the maximum mass of QSs in Fig.~\ref{fig_profiles_vary_lambda}. 
As it is clear from the Fig.~\ref{fig_profiles_vary_lambda} and Table \ref{table2} that  a larger $\bar{\lambda}$ 
leads to a large value of maximum mass, since a larger $\bar{\lambda}$  maps to a stiffer EoS.  According to Table \ref{table2}, the maximum gravitational mass goes upto the  $M_{\rm{max}}$= 3.25 $M_{\odot}$ with radius 13.56 km for $\bar{\lambda}= 1$, while we have recorded the $M_{\rm{max}}$= 2.37 $M_{\odot}$ with radius 11 km for $\bar{\lambda}= 0.1$. 
Observe that all QS sequences reach a maximum mass of at least $M_{\rm{max}} >$ 2 $M_{\odot}$. From our calculation, it is evident that when $\bar{\lambda}> 0.3$ the maximum mass of the corresponding QSs will meet the lower mass limit of the secondary component of GW190814. The results for R-R gravity theory are presented in the solid curves while dashed curves present the Rainbow gravity theory only (i.e., $\eta=1$) as depicted in Fig.~\ref{fig_profiles_vary_lambda}. Next, we present the  $(M-M/R)$ diagram in the lower panel of Fig.~\ref{fig_profiles_vary_lambda} using the same set of parameters. We notice that the central energy density of the maximum mass decreases as the $\bar{\lambda}$ of R-R gravity increases, see Table \ref{table2}. It is seen that the maximum compactness increases as the value of $\bar{\lambda}$ increases and lies within the range of $0.319 < M/R < 0.356$.

\begin{table}
\caption{\label{table2}  Summary of the resulting properties of the isotropic QSs for $B_{\rm eff} = 90$ MeV/fm$^3$, $\Bar{\lambda} \in [0.1,1.0]$, $\Sigma = 1.05$, and $\eta = 0.8$. }
\begin{ruledtabular}
\begin{tabular}{ccccc}
$\Bar{\lambda}$  & $M$ [$M_\odot$]  &  $R_{M}$  [\rm{km}]  & $\rho_c$ [MeV/fm$^3$] & $M/R$  \\
\colrule
0.1 &  2.37  &  11.00  &  1,069  &  0.319  \\
0.5 &  2.89  &  12.56  &  844  &  0.341  \\
1.0 &  3.25  &  13.56  &  732  &  0.356 
\end{tabular}
\end{ruledtabular}
\end{table}

\begin{table}
\caption{\label{table3} 
Summary of the resulting properties of the isotropic QSs for $B_{\rm eff} = 90$ MeV/fm$^3$, $\Bar{\lambda} \in 0.1$, $\Sigma \in [1.00,1.15]$, and $\eta = 0.8$.}
\begin{ruledtabular}
\begin{tabular}{ccccc}
$\Sigma$  & $M$ [$M_\odot$]  &  $R_{M}$  [\rm{km}]  & $\rho_c$ [MeV/fm$^3$] & $M/R$  \\
\colrule
1.00 &  1.92  &  10.10  &  1,238  &  0.281  \\
1.05 &  2.22  &  10.61  &  1,238  &  0.310  \\
1.10 &  2.56  &  11.11  &  1,238  &  0.341  \\
1.15 &  2.92  &  11.62  &  1,238  &  0.372 
\end{tabular}
\end{ruledtabular}
\end{table}

\begin{table}[h]
\caption{\label{table4} 
Summary of the resulting properties of the isotropic QSs for $B_{\rm eff} \in [60,90]$ MeV/fm$^3$, $\Sigma = 1.05$, $\bar{\lambda} = 0.1 $, and $\eta = 0.8$.}
\begin{ruledtabular}
\begin{tabular}{ccccc}
$B_{\rm eff}$  & $M$ &  $R_{M}$ & $\rho_c$ & $M/R$  \\
MeV/fm$^3$ & $M_\odot$ & \rm{km} & MeV/fm$^3$ &  \\
\colrule
60 &  2.90  &  13.43  &  731  &  0.320  \\
70 &  2.68  &  12.45  &  844  &  0.319  \\
80 &  2.51  &  11.66  &  957  &  0.319  \\
90 &  2.37  &  11.00  &  1070  &  0.319 
\end{tabular}
\end{ruledtabular}
\end{table}

\subsection{Profiles for variation of the Rainbow parameter $\Sigma$}

Next, we continue our study of by examining the effect of Rainbow parameter $\Sigma$ on the properties of QSs. 
In Fig.~\ref{fig_profiles_vary_sigma} we focus on the behavior of $(M-R)$ and $(M-M/R)$ relations with varying 
Rainbow parameter $\Sigma \in $ [1.00, 1.15]. In this case the other parameters are  $B_{\rm eff} = 90$ MeV/fm$^3$,  $\bar{\lambda} = 0.1 $ and $\eta = 0.9$, respectively. Some quantities related to the maximum mass of QSs
and its corresponding radius are shown in Table~\ref{table3}. Depending on the model, we see that 
the maximum mass increases monotonically with increasing values of $\Sigma$ and comfortably well above the two solar mass.  Regarding the values shown in Table~\ref{table3}, we remark that the maximum gravitational mass goes upto the  $M_{\rm{max}}$= 2.92 $M_{\odot}$ with radius 
11.62 km, which is much higher than GR counterpart. With these results, it is reasonable to expect a massive QS with $M   \sim 3 M_{\odot}$. Running over the mass and radius ranges, we conclude that the present model is compatible with the gravitational-wave event GW190814. Finally, we move on to the  $(M-M/R)$ diagram in the lower panel of Fig.~\ref{fig_profiles_vary_sigma}. We note that the maximum compactness increases as the value of $\Sigma $ increases and lies within the range of $0.281 < M/R < 0.372$. It is noteworthy that the variation of the $\Sigma$ does not affect the central energy density at the maximum mass at all, see Table~\ref{table3} for detail. 

\subsection{Profiles for variation of the effective bag constant  $B_{\rm eff}$}

In these proceedings we add a new aspect to this discussion is the variation of $B_{\rm eff} \in [60,90]$ MeV/fm$^3$. 
We also adopt $\Sigma = 1.05$, $\bar{\lambda} = 0.1 $ and $\eta = 0.8$ are fixed.  We show our results for the  $(M-R)$ and $(M-M/R)$ relations using the quark matter  EoS (\ref{eos_p}) and presented in Fig.  \ref{fig_profiles_vary_B}. 
We can see from Fig. \ref{fig_profiles_vary_B} that the maximum mass decreases with increasing values of $B_{\rm eff}$.
Still, regarding the values shown in Table \ref{table4}, we remark that the maximum mass of QS is $M_{\rm{max}}$= 2.90 $M_{\odot}$ with radius 13.43 km at $B_{\rm eff}= 60$ MeV/fm$^3$. At the same time,  we gather data for the maximum masses and their corresponding radii for different values 
of $B_{\rm eff}$, and see that the maximum mass exceeding $M_{\rm{max}}>$2 $M_{\odot}$ constraint depending on the model parameters. In this figure, we have included the observational constraints, same as of Fig. \ref{fig_profiles_vary_eta}. Finally, we plot the compactness as function of gravitational mass in the lower panel of Fig. \ref{fig_profiles_vary_B}. For $(M-M/R)$ dependencies presented in Table \ref{table4}, we can say that maximum compactness increases for less interacting quarks, and the maximum compactness can reach values up to 0.320.

\section{The static stability criterion, adiabatic index and the sound velocity}
\label{sec5}

\begin{figure*}
    \centering
    \includegraphics[width = 8 cm]{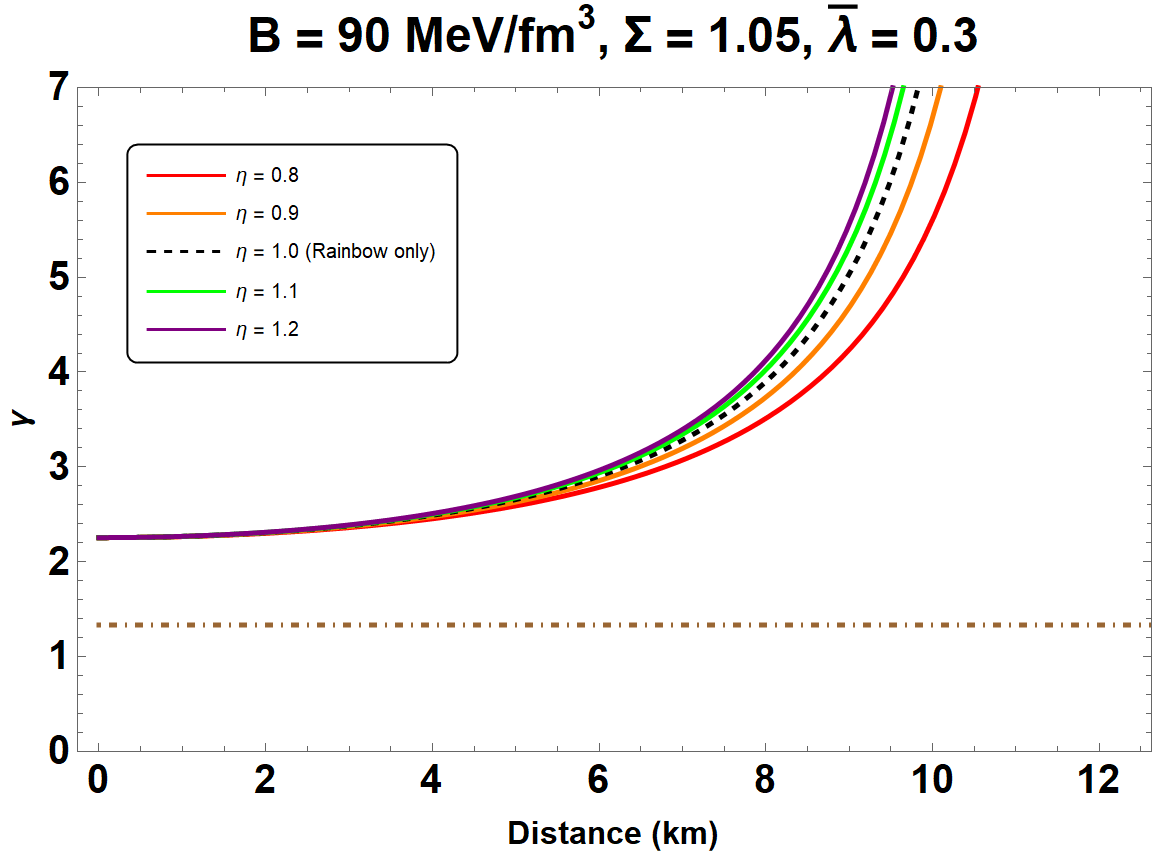}
    \includegraphics[width = 8 cm]{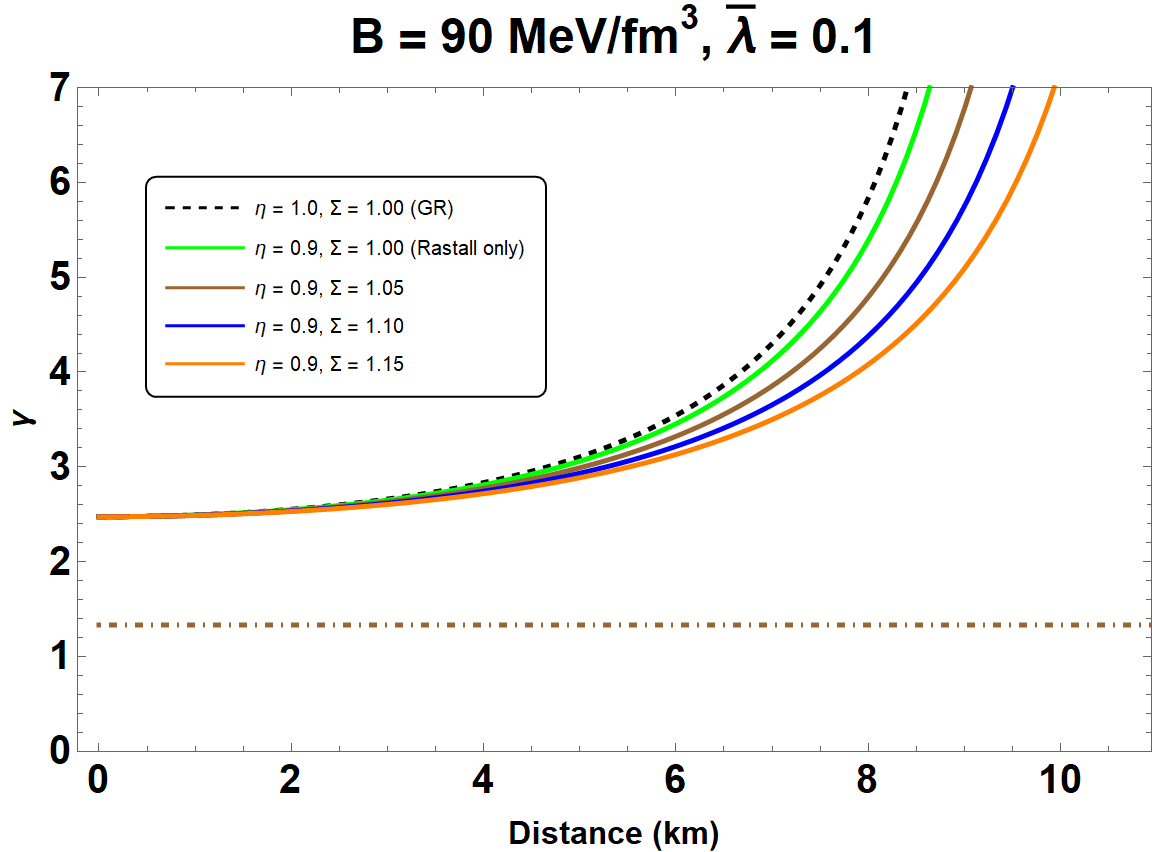}
    \includegraphics[width = 8 cm]{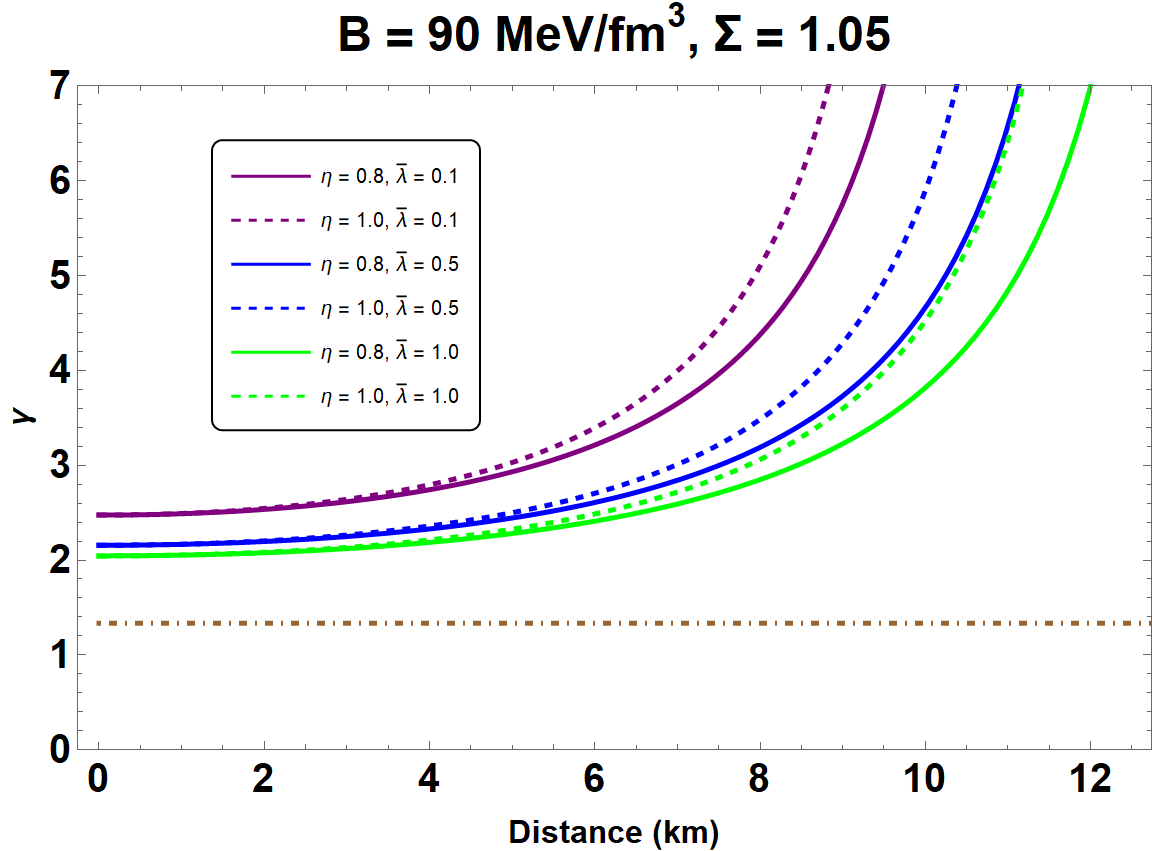}
    \includegraphics[width = 8 cm]{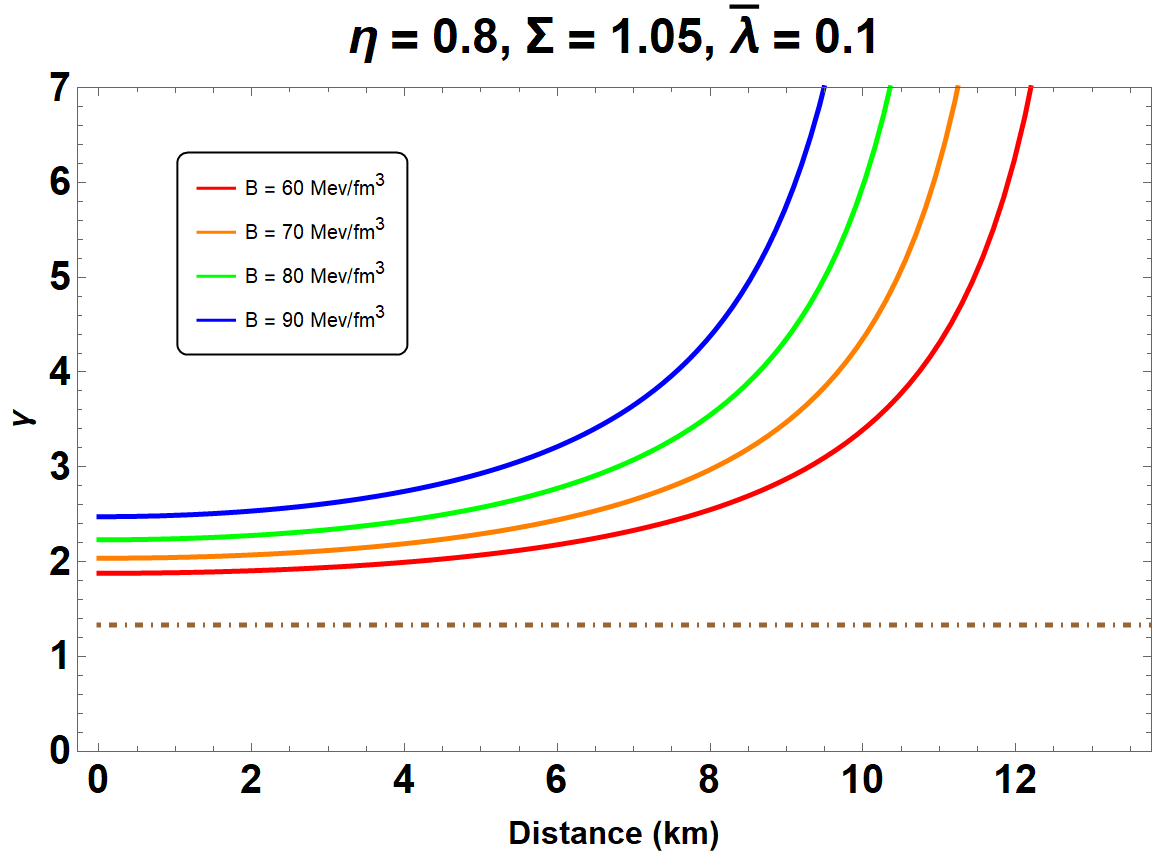}
    \caption{From top to bottom panel, we plot the adiabatic index, $\gamma $,  along the radial distance. Used  parameters are same as of Fig.~\ref{fig_profiles_vary_eta} to Fig.~\ref{fig_profiles_vary_B} with fixed $\rho_c = 562$ MeV/fm$^3$. The lower bound of the adiabatic index  $\Gamma = 4/3$ is presented in the brown dot-dash line. }
    \label{fig_adiabatic}
\end{figure*}

For completeness, we provide here the $(M-\rho_c)$ profiles that are related to the stability of the configuration, 
which is known as the \textit{static stability criterion} \cite{Harrison,ZN}. Through this condition, one can identify the separable region from stable to unstable one at the turning point $(M_{\text{max}}, R_{M_{\text{max}}})$. 
But, it should be noted that this is a necessary condition but not sufficient. Making the ansatz of the static stability criteria \cite{Harrison,ZN}, it states that
\begin{eqnarray}
    &&\frac{d M}{d \rho_c} < 0 \rightarrow \text{unstable configuration}, \\
    &&\frac{d M}{d \rho_c} > 0 \rightarrow \text{stable configuration}, 
\end{eqnarray}
to be satisfied in all configurations. To be more specific, we can say that the stable QSs are found in the region where $dM/d\rho_c >0$.  In Fig. \ref{fig4}, we present a set of all $(M-\rho_c)$ graphs computed for the models proposed in Subsections A, B and C, separately. From the observational point of view, the $(M-\rho_c)$ curves are indistinguishable at the low central density region, whereas at the high central density region the difference between curves is prominent. In Fig. \ref{fig4}, the pink points determine the stable
and unstable configurations against radial oscillations. 

\begin{figure*}
    \centering
    \includegraphics[width = 8.3 cm]{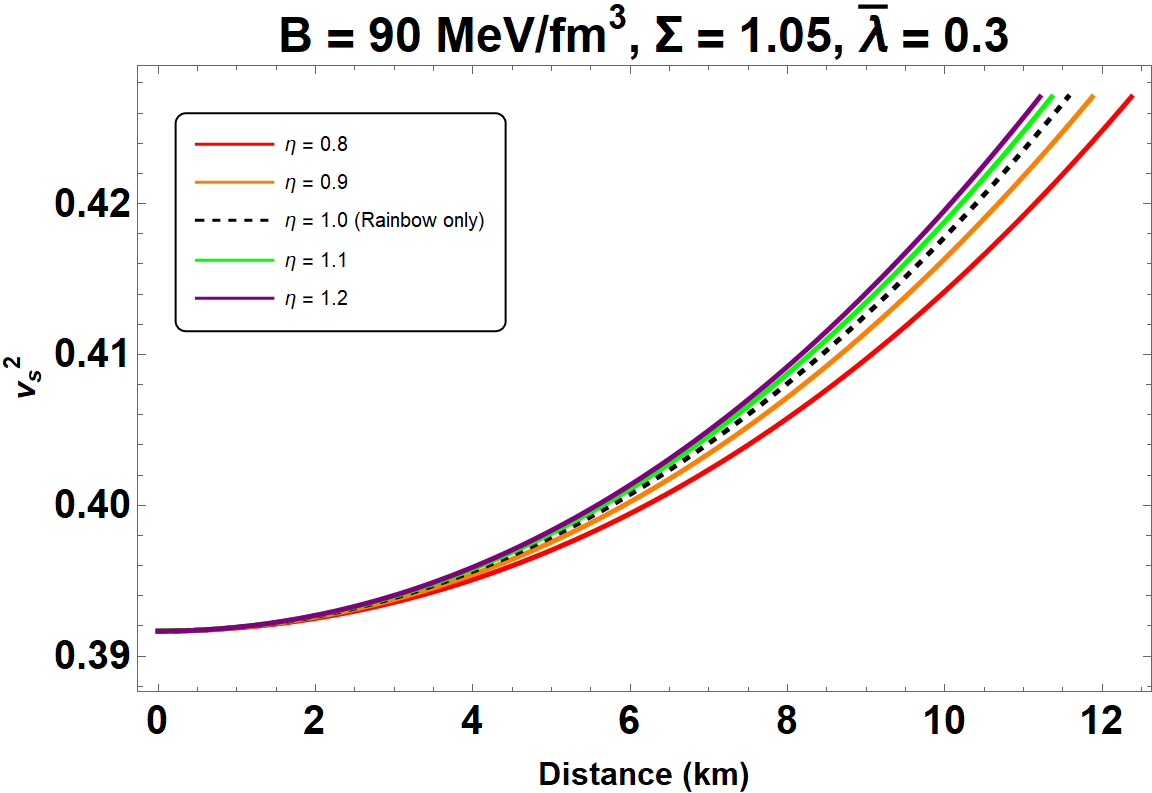}
    \includegraphics[width = 8.3 cm]{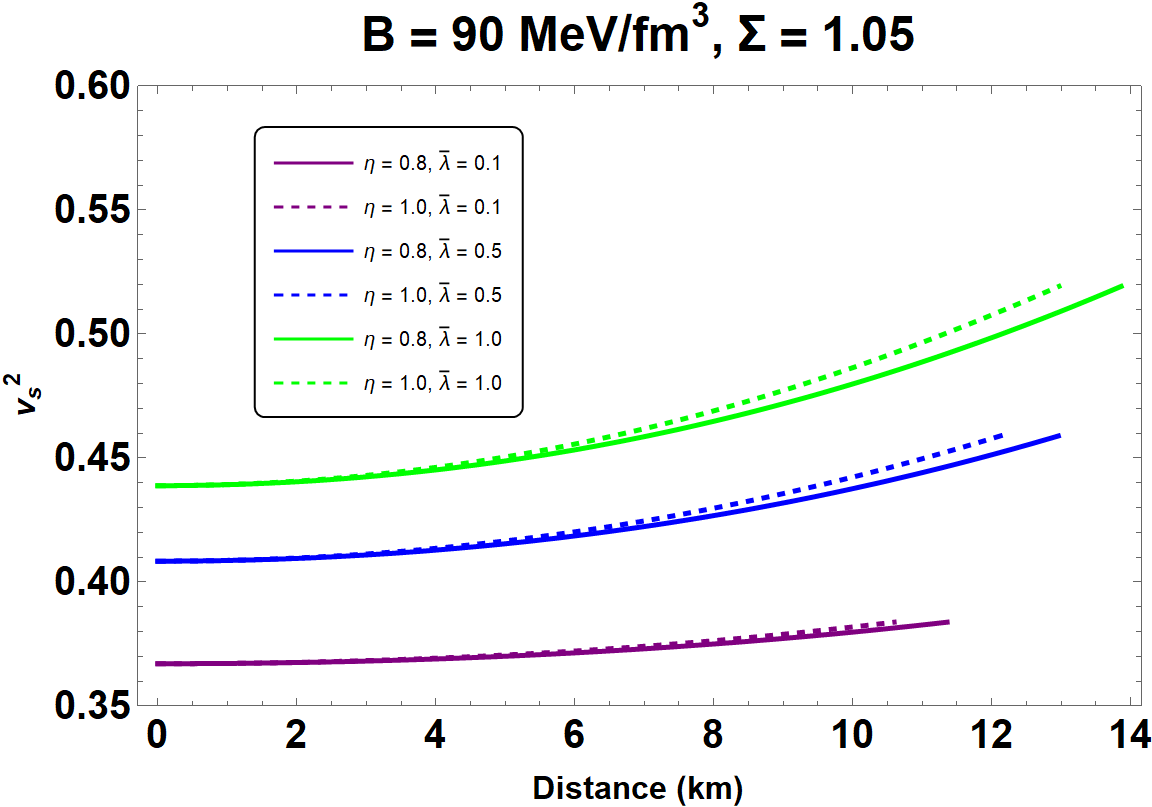}
    \includegraphics[width = 8.3 cm]{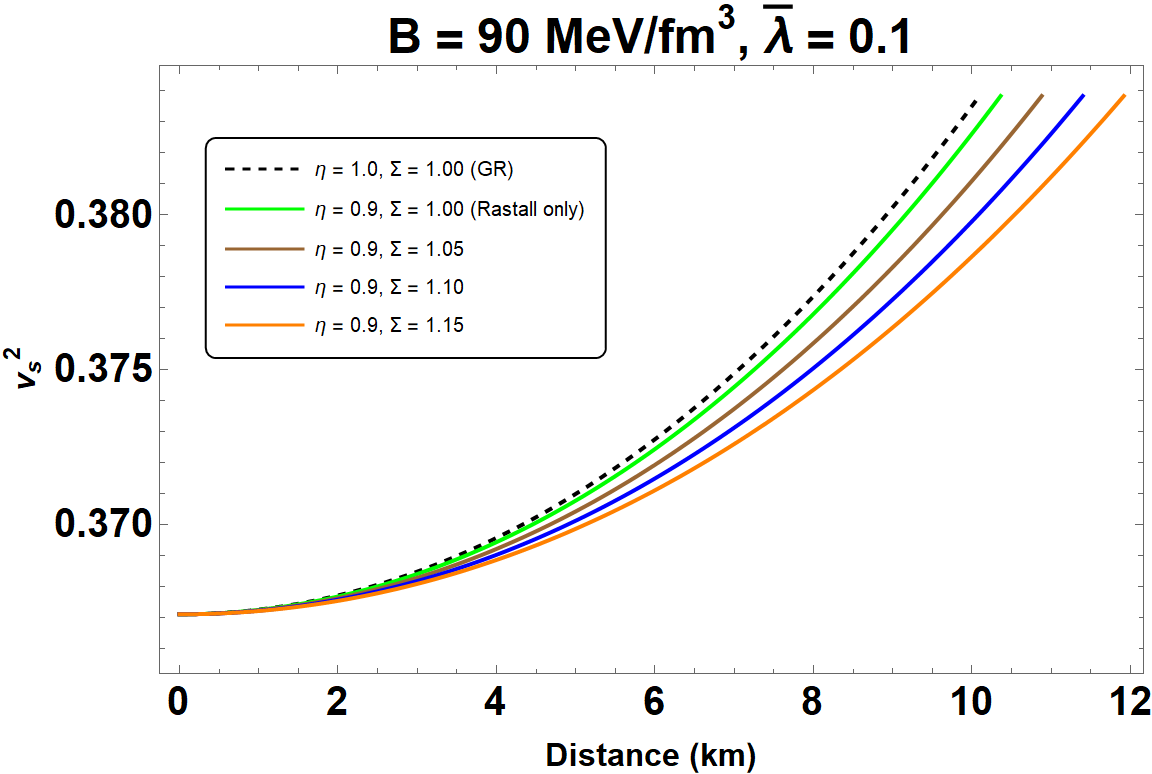}
    \includegraphics[width = 8.3 cm]{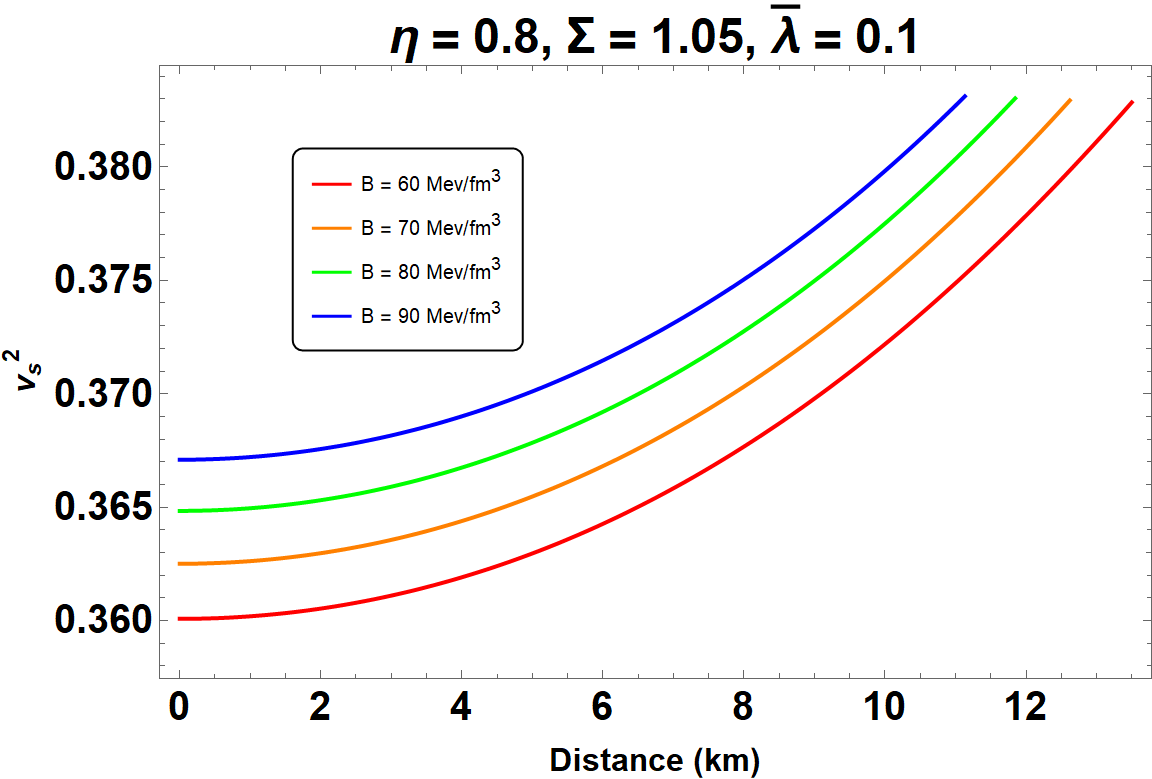}
    \caption{The velocity of sound in QSs in the R-R gravity for the given EoS (\ref{eos_p}). Used  parameters are same as of Fig.~\ref{fig_profiles_vary_eta} to Fig.~\ref{fig_profiles_vary_B} with fixed $\rho_c = 562$ MeV/fm$^3$.}
    \label{fig_SS}
\end{figure*}

We now focus on the dynamical stability of QSs based on the variational method for the given EoS (\ref{eos_p}). This approach was introduced by Chandrasekhar in 1964 \cite{Chandrasekhar}, which can be defined via the speed of sound through
\begin{eqnarray}
    \gamma = \left( 1 + \frac{\rho}{p} \right) \left( \frac{dp}{d\rho} \right), \label{adiabatic_index}
\end{eqnarray}
where $dp/d\rho$ is the square of sound speed and $\gamma$ is the dimensionless, called the  adiabatic index.  Here, we recall certain restrictions on $\gamma$, that determine whether the condition of a stable spherical static object do exists or not. We identify this condition by the critical adiabatic index $\gamma_{\text{cr}} = 4/3$. Below this value the configuration is unstable against radial perturbations \cite{Glass}. In Fig.~\ref{fig_adiabatic}, the adiabatic index $\gamma$ has been plotted for three consecutive cases studied in Subsections A, B and C, respectively. Observing the
 Fig.~\ref{fig_adiabatic}, it can be ensured that the adiabatic index increases along the
radial distance, and exceeds the lower bound of $\gamma_{\text{cr}}$. This leads to the stability of QSs.

We end this section by studying the sound speed. This is another indicator related to the stability of compact objects, and defined by $ v^2_s = dp/ d\rho $.  Since, we know that 
for a physically reasonable model, it is required that the sound speed does not exceed the speed of light, i.e. in our units $ v^2_s <1$. Using the Eq. (\ref{eos_p}), we plot the sound speed as a function of radial distance for a fixed value of the central pressure fixed $\rho_c = 562$ MeV/fm$^3$ in Fig. \ref{fig_SS}. As evident from the figures, the requirements for sound speed are satisfied throughout the stellar interior for all the studied cases.  

\section{Conclusions} \label{sec6}

Quantum chromodynamics (QCD) is the theory of strong interactions between quarks and gluons. Since, QCD has been studied for decades, but not completely clear to us still now. Concerning this, a recent study \cite{Zhang:2020jmb} demonstrated effects from QCD
interactions such as color-superconductivity and perturbative QCD (pQCD) corrections that lead to a new EoS called 
interacting quark matter (IQM). These corrections to the EoS may reveal new physical phenomena in  strongly interacting regime which may be found in the core of compact objects. The present article aims to explore the properties of stable compact stars made of IQM in R-R gravity theory. The R-R theory is a newly proposed
modified theory of gravitation constructed by combining two distinct theories, namely, the Rastall theory and the gravity's rainbow formalism. 

Summing up, in this work we solved numerically the modified TOV equations (\ref{master1}) and (\ref{master3}), and examined the diagrams related to ($M-R$), ($M-\rho_c$) and ($M-M/R$) for all the considered cases. We have separately studied the effects of ($\Sigma, \eta, \Bar{\lambda}, B_{\rm eff}$) parameters on the properties of static QSs. The resulting numerical values for the masses and radii of the QSs are compatible with data from various observed pulsars including constraints from GW190814 and GW170817 events in all the studied cases. It is also to be noted that by increasing values of $\bar{\lambda}$, the maximum mass of QS increases and comfortably exceeds 2$M_{\odot}$. Moreover, we show the
possibility of achieving high masses like $M> 3M_{\odot}$ or more with $R \sim 10-14$ km in modified gravity.
 One may note that a similar conclusion is supported by several recent works in different modified gravity frameworks \cite{Bora:2022qwe, Bora:2023zhp, Pretel:2023avv, Bhar:2023yrf}. In Ref. \cite{Pretel:2023avv}, authors showed that the model parameter in momentum squared gravity also plays a significant role in this. They discovered that as the model parameter $\alpha$ increases, the maximum mass initially rises, reaching a peak value before decreasing.

Finally, we comment on the stability of QSs based on the static stability criterion, adiabatic index and the sound velocity. The results of our findings are interesting since the stellar stability has been confirmed by performing those analyses. The stability analysis against adiabatic radial oscillations for QSs
in R-R gravity will be left for a future work.

\begin{acknowledgments}
T. Tangphati is financially supported by Research and Innovation Institute of Excellence, Walailak University, Thailand under a contract No. WU66267. A. Pradhan thanks to IUCCA, Pune, India for providing facilities under associateship programmes.
\end{acknowledgments}\


\end{document}